\documentclass[lettersize,journal]{IEEEtran} 

\usepackage{booktabs}
\usepackage{balance}
\usepackage[mathscr]{euscript}
\usepackage{verbatim}
\usepackage{amsmath}
\usepackage{amssymb}
\usepackage{color}
\usepackage{graphicx}
\usepackage{amsbsy}
\usepackage{bm}
\usepackage{epsfig}
\usepackage{amsthm}
\usepackage{fixltx2e}
\usepackage{xcolor}
\usepackage{dblfloatfix}
\usepackage{bbm}
\usepackage[flushleft]{threeparttable}
\usepackage{stmaryrd}
\usepackage{multirow,url}
\usepackage{algorithm}
\usepackage{algorithmic}

\usepackage{subfigure}

\usepackage{cite}

\allowdisplaybreaks[3]

\newcommand{\be}{\begin{equation}}
\newcommand{\ee}{\end{equation}}
\newcommand{\bea}{\begin{eqnarray}}
\newcommand{\eea}{\end{eqnarray}}

\newcommand{\bsm}{\begin{small}}
\newcommand{\esm}{\end{small}}

\newcommand{\bfa}{\mathbf{a}}

\newcommand{\hcomm}{\mathbf{h}_{i,n,k}}

\newcommand{\hstack}{\mathbf{h}_{i,k}}

\usepackage{etoolbox}
\makeatletter 
\pretocmd\@bibitem{\color{black}\csname keycolor#1\endcsname}{}{\fail}
\newcommand\citecolor[1]{\@namedef{keycolor#1}{\color{blue}}}
\makeatother

\begin{document}

\title{Cooperative  ISAC for Joint Localization and Velocity Estimation in Cell-Free MIMO Systems
\thanks{
Manuscript received on Feb. 13, 2025; revised on Jul. 15, 2025; accepted on Sept. 6, 2025. The work of Zihuan Wang and Vincent Wong was supported in part by the Government of Canada Innovation for Defence Excellence and Security (IDEaS) program and the Digital Research Alliance of Canada (alliancecan.ca). Robert Schober’s work was supported by the Federal Ministry for Research, Technology and Space (BMFTR) in Germany in the program of ``Souver\"{a}n. Digital. Vernetzt.'' joint project 6G-RIC (Project-ID 16KISK023), the Deutsche Forschungsgemeinschaft (DFG, German Research Foundation) under projects SFB 1483 (Project-ID 442419336, EmpkinS) and Horizon Europe Marie Skodowska-Curie Actions (MSCA)-UNITE under project 101129618.

This paper has been published in part in the {\it Proceedings of   IEEE International Conference on Acoustics, Speech, and Signal Processing (ICASSP)}, Apr. 2025 \cite{ZW-ICASSP}. (\textit{Corresponding author: Vincent W.S. Wong})}
}

\author{Zihuan Wang,~\IEEEmembership{Member,~IEEE,} Vincent W.S. Wong,~\IEEEmembership{Fellow,~IEEE,} and Robert Schober,~\IEEEmembership{Fellow,~IEEE}
         \thanks{Zihuan Wang and Vincent W.S. Wong are with the Department of Electrical and Computer Engineering, The University of British Columbia, Vancouver, Canada (e-mail: \{zihuanwang, vincentw\}@ece.ubc.ca). Robert Schober is with the Institute for Digital Communications, Friedrich-Alexander University of Erlangen-Nuremberg, Erlangen 91058, Germany (email: robert.schober@fau.de).
         
         Color versions of one or more of the figures in this paper are available online at https://ieeexplore.ieee.org.
         }
}  
         
\pagestyle{empty}

\maketitle

\thispagestyle{empty}

\begin{abstract}
In this paper, we explore a cooperative integrated sensing and communication (ISAC) framework that utilizes orthogonal frequency division multiplexing (OFDM) waveforms. Under the control of a central processing unit (CPU), multiple access points (APs) collaboratively perform multistatic sensing while providing communication service in a cell-free multiple-input multiple-output (MIMO) system. Achieving high sensing accuracy requires the collection of global sensing information at the CPU, which can lead to significant fronthaul signaling overhead due to the feedback of the sensing signals from each AP. To tackle this issue, we propose a collaborative processing scheme in which the APs locally compress and quantize the received sensing signals before forwarding them to the CPU. The CPU then aggregates the information from all APs to estimate the location and velocity of the targets. We develop a distributed vector-quantized variational autoencoder (D-VQVAE) to enable an end-to-end implementation of this scheme. D-VQVAE  consists of distributed encoders at the APs to locally encode the received sensing signals, codebooks for quantizing the encoded results, and a decoder at the CPU for location and velocity estimation. It effectively reduces the amount of data transmitted from each AP to the CPU while maintaining a high sensing accuracy. 
We employ a collaborative learning-assisted scheme to train D-VQVAE in an end-to-end manner. 
Simulation results show that the proposed D-VQVAE network outperforms the baseline schemes in sensing accuracy and reduces fronthaul signaling overhead by $99 \%$ when compared with the centralized sensing approach.
\end{abstract}


\vspace{-0.0 cm}
\section{Introduction}
\label{sec:intro}
\vspace{-0.0 cm}

The sixth generation (6G) wireless networks aim to support emerging services such as immersive extended reality, intelligent transportation systems, and smart manufacturing, all of which demand increased spectrum usage for both sensing and communication. As communication systems transition to higher frequency bands, which are conventionally used in radar systems, researchers have been developing algorithms that enable spectrum sharing between sensing and communication systems. Increasing attention is also being given to hardware sharing between both systems in order to reduce the device size, weight, power consumption, and cost. Since the radio frequency (RF) frontend architectures employed by radar sensing and wireless communication systems are similar, the integration of these two functionalities is feasible. 
This overlap in frequency usage and hardware design between radar and communication systems has driven the emergence of integrated sensing and communication (ISAC).

ISAC has been identified as a key use case for 6G and can benefit various applications, such as the Internet of Things
(IoT) \cite{IoTMag} and vehicular networks \cite{ZW-JSTSP}. It enables wireless networks to simultaneously transmit information and receive sensing echoes through a unified infrastructure and shared resources, thus improving both spectrum and energy efficiencies \cite{FL-JSAC22,Nuria}. 
Orthogonal frequency division multiplexing (OFDM) is a widely used waveform in ISAC systems. OFDM waveforms can effectively combat frequency-selective fading and provide high data rates \cite{JSTSP_Zhang}. Moreover, OFDM waveforms exhibit Doppler tolerance and do not suffer from range-Doppler coupling, which makes them suitable for radar sensing \cite{SHD,JBS-TVT}. Unlike the conventional OFDM radar systems, where the transmit signals do not carry useful information, the OFDM signals used in ISAC systems contain modulated data for communication, which introduces different phase shifts in the transmit signals. These phase shifts must be considered for sensing \cite{YW-TCOM,YL-TVT}. By analyzing the reflected sensing signals, the range and velocity of the targets can be estimated. The use of multiple-input multiple-output (MIMO) architectures further provides additional degrees of freedom (DoFs) in the spatial domain, enabling the extraction of angle, range, and velocity information of targets from the reflected sensing signals  \cite{MAI-ICC,ZX-TSP,ZXiao-TSP}.

However, the conventional monostatic ISAC configuration with a single MIMO base station (BS) faces several limitations. On one hand, the communication performance may be degraded due to inter-cell interference, and users at the cell edge may experience poor service. On the other hand, relying on a single BS limits spatial diversity, which makes accurate sensing in complex environments with multiple targets more challenging.
To address these issues, cooperative ISAC exploiting cell-free MIMO architectures has been proposed \cite{LX-IWC23,ZZMNET,UD-TCOM,ZW-ISACom}, where multiple access points (APs) are distributed across the coverage area to jointly provide seamless communication service and gather multiview sensing observations. These APs are connected to a central processing unit (CPU), enabling effective collaboration among the APs. Cooperative ISAC can enhance communication through coordinated multipoint transmission, providing more reliable connections and higher throughput. The multistatic sensing enabled by cooperative ISAC offers wider sensing ranges and captures multiview sensing information, leading to increased sensing accuracy.


\subsection{Related Work on ISAC}

Researchers have explored the use of OFDM waveforms in ISAC systems. In \cite{YW-TCOM}, the OFDM waveform is applied for range and velocity estimation without compromising the communication performance. 
The sensing channel frequency response is first estimated, where the communication information is removed. A deep learning-based algorithm is then applied to extract the range and velocity information from the estimated frequency response. 
In \cite{YL-TVT}, a super-resolution method is proposed for range and velocity estimation by exploiting the translational invariance of the received sensing signals in frequency and time domain. 
These studies focus on single-input single-output (SISO) scenarios, whereas in practice MIMO architectures are widely employed at BSs to enhance the spectral efficiency through beamforming, enabling the extraction of angle, range, and velocity information from the reflected sensing signals \cite{ZX-TSP,MAI-ICC,ZXiao-TSP}. 
In \cite{MAI-ICC}, sensing parameter estimation using ISAC is studied. The multiple signal classification (MUSIC) algorithm is used for angle estimation based on the sensing signals, followed by the extraction of delay and Doppler shifts using a two-dimensional (2D) discrete Fourier transform (DFT).
In \cite{ZX-TSP}, the angle, range, and velocity of targets are sequentially extracted from the received sensing signals using DFT. In \cite{ZXiao-TSP}, the sensing parameters are jointly estimated through spectral analysis across the space, frequency, and time domains. 
Uplink sensing is studied in \cite{HL-TCCN25}, where the initial delay is estimated and refined iteratively, followed by angle and Doppler estimation.
Given the estimated angle and range, the target locations can be determined from the corresponding geometric relationships.

The aforementioned works focus on a single BS for monostatic sensing, which may result in limited sensing performance due to restricted spatial diversity. Moreover, the velocity obtained based on these approaches is only the radial component deduced from Doppler shifts, while the tangential component of the velocity is unavailable. Thus, the complete velocity vector of targets cannot be obtained. 
To overcome these limitations, cooperative ISAC has been proposed, utilizing geographically distributed APs to gather multiview sensing information.  
In \cite{FA-JCS}, the angular information of the targets in a cell-free ISAC system is jointly estimated by distributed APs using deep neural networks (DNNs).  
In \cite{WJ-TWC}, an iterative angle estimation scheme is proposed, where the angle is refined through iterative coarse and fine estimation procedures. In \cite{ZL-TWC24}, a cooperative target localization scheme is proposed, where the angle and range information is first extracted from the received signals at each AP, followed by selecting a set of APs with high correlations for cooperative localization. In \cite{MSH-WCNC24}, a maximum likelihood estimation based target localization scheme is proposed, where the APs collaboratively estimate the locations of the targets using the transmit data payload from the uplink. 
In \cite{QS-JSAC},  a two-phase scheme is
proposed for target localization by using cooperative ISAC. In \cite{zhang-TCOM}, ranges are first estimated based on a two-dimensional fast Fourier transform (2D-FFT)-based algorithm. Then, the target locations are estimated based on the range measurements.  In \cite{ML-TAES20}, cooperative ISAC in cell-free MIMO systems is explored, where each AP employs a compressive sensing (CS)-based algorithm to estimate the range, angle, and radial velocity of the targets. Then, the CPU determines the targets’ locations and velocities by leveraging the geometric relationships. In \cite{ZW-TVT24}, cooperative ISAC for target sensing based on symbol-level information is investigated. The APs preprocess the collected symbols to extract the state parameters and phase features of the target. The extracted information is then fused at the CPU for estimating the location and velocity of the target. In \cite{PL-ICL}, collaborative motion recognition based on distributed ISAC is studied. A federated edge learning based scheme is proposed for
collaborative recognition while preserving data privacy.

We note that most of the existing works adopt a two-phase strategy for target localization and velocity estimation, which can lead to error propagation and potential performance degradation due to inaccuracies in the estimated range, angle, and radial velocity. In \cite{ZW-ICASSP}, which is the conference version of this paper, we proposed a DNN consisting of convolutional neural network (CNN) layers to encode the reflected sensing echoes and directly estimate the location and velocity of the targets at the CPU. This approach bypasses the intermediate step of estimating the sensing parameters (i.e., angle, range, radial velocity) and improves the sensing performance. However, it requires each AP to transmit high-dimensional sensing signals to the CPU for centralized processing, resulting in significant fronthaul signaling overhead.

\subsection{Related Work on Deep Learning for Signal Compression}

Deep learning plays an important role in signal compression in communication systems \cite{ZQ-IWC19}, which aims to reduce the signaling overhead for signal feedback from one node to another. One popular deep learning based approach for signal compression and feedback is the use of autoencoder architectures \cite{YZ-WCL21,MBM-TWC21}. In these works, an encoder is utilized by the users to compress the channel state information (CSI) and generate latent representations with lower dimensions. The compressed CSI is then fed back to the BS, where a decoder reconstructs the original CSI.  Autoencoder-based CSI feedback has demonstrated significant improvement in reconstruction accuracy compared with conventional CS-based methods. Moreover, variational autoencoders (VAEs) have been proposed to tackle signal compression \cite{JG-TCOM22}. VAEs model the latent space probabilistically by learning a distribution over the latent variables. This allows for more robust representations that can be beneficial in highly dynamic or noisy environments. In VAE, the encoder compresses the CSI by generating a distribution (typically Gaussian) over the latent space, from which samples are drawn and transmitted to the BS. The decoder then reconstructs the CSI from these samples. More recently, vector quantized variational autoencoders (VQVAEs) have been utilized due to their ability to produce discrete latent codes via codebook-based quantization \cite{XS-ICL24,JS-WCL24}. In VQVAEs, the encoder output is mapped to the nearest codebook entry, resulting in compact index-based representations that are well suited for quantized digital feedback.
Furthermore, large foundation models have been developed for wireless applications. These models leverage transformer models with multi-head attention mechanisms to capture complex spatial and temporal relationships in wireless channels and aim to achieve generalizations across different tasks in wireless systems \cite{SA-arXiv25}. By pre-training on large-scale data, large foundation models can serve as a universal feature extractor for various tasks in wireless systems. 

\subsection{Motivations and Contributions}

In this paper, we consider a cooperative ISAC framework for multistatic sensing in cell-free MIMO systems. Under this framework, a set of distributed transmit APs send information-carrying OFDM signals to the communication users. These signals also reach the sensing targets within the area of interest, generating sensing echoes. These echo signals are then collected by a separate set of distributed receive APs.
From the existing works, we observe that there are typically two approaches for target localization and velocity estimation within the cooperative ISAC framework. The first approach is \textit{fully distributed sensing} \cite{zhang-TCOM}, where the location and velocity of the targets are estimated in two phases. Each receive AP first independently estimates the angle, range, and radial velocity of each target, and then forwards these parameters to the CPU. Based on the received parameters, the CPU determines the location and velocity of the targets. This approach incurs a small signaling overhead, as the estimated sensing parameters can be represented by a small number of bits. However, it may result in poor sensing accuracy due to estimation error propagation. The second approach is \textit{centralized sensing} \cite{LX-IWC23}, where the receive APs send the raw sensing signals to the CPU. The CPU estimates the location and velocity of the targets based on the received sensing signals. While this approach can improve the sensing accuracy by leveraging global information, it suffers from significant fronthaul signaling overhead for transmitting the raw sensing signals.

To address the aforementioned issue, in this paper, we propose to split the entire sensing process between the receive APs and the CPU, which enables signal preprocessing to be performed locally at the receive APs. Our proposed approach reduces the amount of data transmitted over the fronthaul links while ensuring that useful sensing information is obtained by the CPU. The CPU can effectively perform target sensing by fusing the information obtained from all the receive APs, thereby providing a high sensing accuracy.
The main contributions of this paper are summarized below:

\setlength{\hangindent}{2.0em}
$\bullet$ 
To ensure that the CPU can collect sufficiently accurate global sensing information while incurring a low signaling overhead over the fronthaul links, we develop a collaborative processing scheme. In this scheme, signal preprocessing is performed locally at each receive AP to compress the sensing signals and extract sensing-related features, followed by quantization. The receive APs then send the quantized results to the CPU. Thus,  the amount of data transmitted over the fronthaul links is reduced. The CPU fuses the information obtained from all the receive APs to estimate the location and velocity of the targets. Compared with fully distributed and centralized sensing approaches, our proposed scheme offers a trade-off between signaling overhead and sensing performance.

\setlength{\hangindent}{2.0em}
$\bullet$ We propose a distributed vector-quantized variational autoencoder (D-VQVAE) for collaborative processing, which is re-architected from the original VQVAE and tailored for cooperative ISAC in cell-free MIMO systems. The D-VQVAE network comprises distributed encoders and codebooks at the receive APs and a decoder at the CPU. The receive APs encode the reflected sensing signals locally through their respective encoders. Then, the encoded continuous latent representations are quantized into discrete latent feature vectors based on a codebook, where only the indices of these vectors are transmitted to the CPU. Finally, the CPU employs a decoder to estimate the location and velocity of the targets based on the information obtained from the receive APs.

\setlength{\hangindent}{2.0em}
$\bullet$ We propose a collaborative learning-assisted framework for end-to-end training of the D-VQVAE network, in which the receive APs and the CPU jointly optimize the encoders, codebooks, and decoder by exchanging intermediate feature representations and gradients.
The mean squared error (MSE) between the network estimates and ground truth serves as the estimation loss to update the encoders and the decoder. Codebook entries are refined via the exponential moving average (EMA) scheme \cite{vqvae}. A commitment loss is employed to encourage the encoded continuous representations to remain close to their assigned codewords and ensures convergence. 

\setlength{\hangindent}{2.0em}
$\bullet$ We conduct simulations and compare our proposed D-VQVAE network with five baseline schemes, including monostatic sensing at a single BS proposed in \cite{ZXiao-TSP}, a CS-based fully distributed sensing scheme \cite{ML-TAES20}, a MUSIC-based distributed sensing extended from \cite{MAI-ICC}, a CNN-based centralized sensing scheme developed in our preliminary work \cite{ZW-ICASSP}, and a distributed variational autoencoder (D-VAE)-based scheme used for an ablation study. We also include the Cram\'er-Rao lower bound (CRLB) for location and velocity estimation to serve as a performance benchmark. Simulation results demonstrate the benefits of cooperative ISAC-assisted target sensing over monostatic sensing. 
The results also show the performance gains of the proposed D-VQVAE network over the baseline schemes while incurring a low fronthaul signaling overhead.

\vspace{-0.cm}
\subsection{Paper Structure and Notations}

The rest of this paper is organized as follows. The system model for cooperative ISAC in cell-free MIMO systems is introduced in Section \ref{sec:system model}. The proposed D-VQVAE network is presented in Section \ref{sec:vqvae}. The collaborative learning-assisted training is described in Section \ref{sec:learning}. The performance evaluation and comparison are provided in Section \ref{sec:performance}. Finally, conclusions are drawn in Section \ref{sec:conclusion}. 

{\it Notations:} We use boldface lower case letters and boldface upper case letters to denote vectors and matrices/tensors, respectively. $(\cdot)^*$, $(\cdot)^{\mathrm{T}}$, and $(\cdot)^{\mathrm{H}}$ are used to denote the conjugate, transpose, and conjugate transpose of a vector or matrix, respectively. $\mathbb{C}^{N}$ and $\mathbb{R}^{N}$ denote the sets of $N$-dimensional vectors with complex entries and real entries, respectively. $\mathcal{CN}(\bm\mu, \bm\Sigma)$ denotes the complex Gaussian distribution, where $\bm\mu$ and $\bm\Sigma$ are the mean vector and covariance matrix, respectively. $\mathbf{I}_N$ indicates an identity matrix of size $N$. $\mathbf{a}[m:n]$ denotes the elements ranging from the $m$-th element to the $n$-th element of vector $\mathbf{a}$.
We use $j$ to denote the imaginary unit which satisfies  $j^2 = -1$.  $\mathrm{Re}\{\cdot\}$ and $\mathrm{Im}\{\cdot\}$ extract the real part and imaginary part of a complex number, respectively.  $\mathbb{E}\{\cdot\}$  denotes the expected value of a random variable. $\mathrm{diag}(\mathbf{a})$ converts a vector $\mathbf{a}$ to a diagonal matrix with the elements of $\mathbf{a}$ on the main diagonal. 
Finally, $\|\cdot\|_2$ and $\| \cdot \|_{F}$ denote the norm of a vector and the Frobenius
norm of a matrix, respectively. 

\vspace{-0. cm}
\section{Cooperative ISAC in Cell-Free MIMO
\label{sec:system model}}

\begin{figure}[t]
\centering
\includegraphics[width=3.0 in]{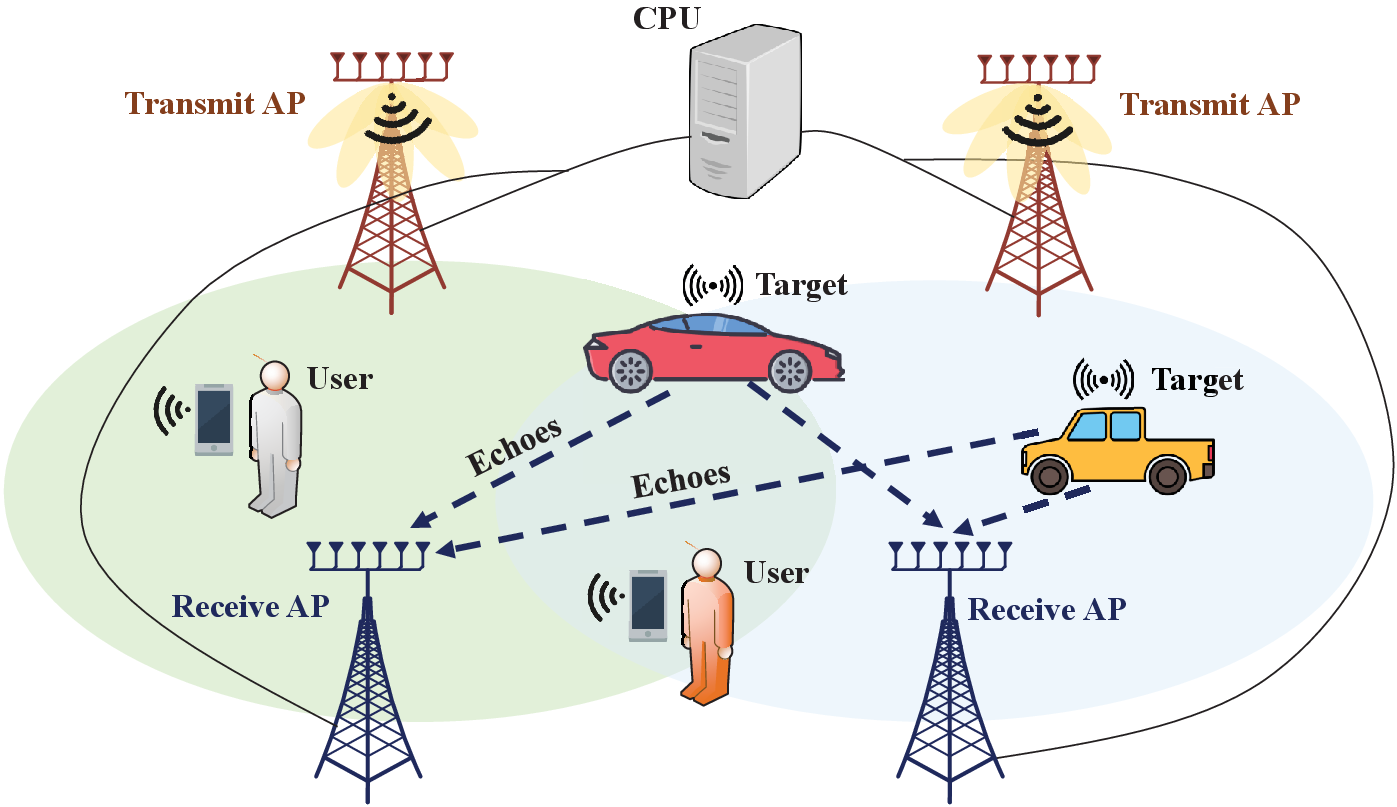}

\caption{Illustration of cooperative ISAC for target sensing in a cell-free MIMO system. The transmit APs send OFDM signals to multiple users for communication, and these signals are reflected by the targets. The reflected sensing signals are collected by the receive APs.}
\label{fig:system model}
\vspace{-0. cm}
\end{figure} 

Consider a cell-free MIMO system with $N$ transmit APs and $M$ receive APs for cooperative ISAC operation. Each transmit AP is equipped with $N_{\mathrm{t}}$ antennas and each receive AP has $M_{\mathrm{r}}$ antennas. 
All APs are connected to a CPU via fronthaul links and they are fully synchronized. There are $K$ single-antenna users, which receive communication signals from the transmit APs, and $Q$ point-like targets to be sensed in the area of interest. 
The transmit APs send OFDM signals to all $K$ users for communication. These transmit signals also reach the targets within the area of interest, generating sensing echoes which are collected by the receive APs. The system model is shown in Fig. \ref{fig:system model}. 
Considering a 2D $(\mathrm{x},\mathrm{y})$ coordinate system, we define $\mathbf{t}_{n} = (t_n^{\mathrm{x}}, t_n^{\mathrm{y}}) \in \mathbb{R}^2$, $n = 1, \ldots, N$, as the location vector of the $n$-th transmit AP. Similarly, let $\mathbf{r}_{m} = (r_m^{\mathrm{x}}, r_m^{\mathrm{y}}) \in \mathbb{R}^2$, $m = 1, \ldots, M$,  denote the location vector of the $m$-th receive AP. We aim to estimate the unknown locations $\mathbf{g}_q=(g_q^{\mathrm{x}}, g_q^{\mathrm{y}}) \in \mathbb{R}^2$ of the targets and their associated velocities along the $\mathrm{x}$- and $\mathrm{y}$-axes, i.e., $\mathbf{v}_q = (v^{\mathrm{x}}_q, v^{\mathrm{y}}_q) \in \mathbb{R}^2$ for $q = 1, \ldots, Q$. We define $\mathbf{g} = [\mathbf{g}^\mathrm{T}_1~ \cdots ~ \mathbf{g}^\mathrm{T}_Q]^\mathrm{T} \in \mathbb{R}^{2Q}$ and $\mathbf{v} = [\mathbf{v}^\mathrm{T}_1~ \cdots ~ \mathbf{v}^\mathrm{T}_Q]^\mathrm{T} \in \mathbb{R}^{2Q}$. Let $\bm\psi = [\mathbf{g}^\mathrm{T} ~ \mathbf{v}^\mathrm{T}]^\mathrm{T} \in \mathbb{R}^{4Q}$ collect the location and velocity vectors of all targets in the area of interest, which needs to be estimated by the CPU.

\begin{figure}[t]
\centering
\subfigure[]{
\includegraphics[width=1.66 in]{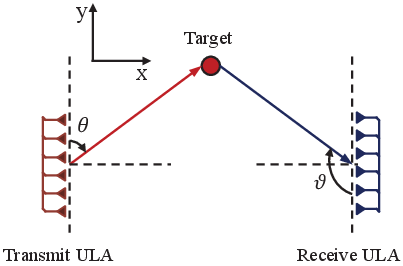}}
\subfigure[]{
\includegraphics[width=1.66 in]{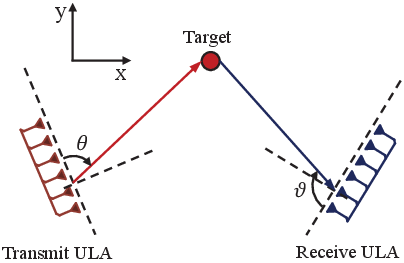}}

\caption{AoD $\theta$ and AoA $\vartheta$ with respect to (w.r.t.) a point-like target. Two different orientations of the transmit and receive ULAs are shown in (a) and (b). In (a), the transmit and receive ULAs are aligned in opposite directions. In (b), the transmit and receive ULAs are oriented with a phase shift between $90^{\circ}$ and $180^{\circ}$ relative to each other.}
\label{fig:coordinate}
\vspace{-0. cm}
\end{figure}

The transmit and receive APs are equipped with uniform linear arrays (ULAs). 
The transmit and receive beam steering vectors are respectively given by
\bea
&&\hspace{-1 cm}\mathbf{a}_{\mathrm{t}}(\theta) = \frac{1}{\sqrt{N_{\mathrm{t}}}} \big[1~ e^{-j2\pi d_{\mathrm{t}}/\lambda_{\mathrm{c}}\cos\theta}\nonumber \\
&&\hspace{1.8 cm}
\cdots ~ e^{-j2\pi(N_{\mathrm{t}}-1)d_{\mathrm{t}}/\lambda_{\mathrm{c}}\cos\theta} \big]^\mathrm{T}, \\
&&\hspace{-1 cm}\mathbf{a}_{\mathrm{r}}(\vartheta) = \frac{1}{\sqrt{M_{\mathrm{r}}}} \big[1~ e^{-j2\pi d_{\mathrm{r}}/\lambda_{\mathrm{c}}\cos\vartheta}\nonumber \\
&&\hspace{1.8 cm}
\cdots ~ e^{-j2\pi(M_{\mathrm{r}}-1)d_{\mathrm{r}}/\lambda_{\mathrm{c}}\cos\vartheta} \big]^\mathrm{T},
\eea 
where $\theta$ and $\vartheta$  denote the angle of departure (AoD) and angle of arrival (AoA), respectively. The definitions of AoD and AoA are illustrated in Fig. \ref{fig:coordinate}, where two different array orientations of the transmit and receive ULAs are presented. We assume that the orientations of the transmit and receive ULAs are known by the CPU. 
Moreover, $d_{\mathrm{t}}$ and $d_{\mathrm{r}}$ denote the transmit antenna spacing and receive antenna spacing, respectively.  $\lambda_{\mathrm{c}} = c/f_{\mathrm{c}}$ represents the wavelength, where $c$ denotes the speed of light and $f_{\mathrm{c}}$ is the carrier frequency.

\subsection{Signal Model}

\begin{figure*}[b]
\hrule
\be \tag{6}
\mathbf{y}_{i,m} [t] = \sum_{n=1}^N 
\underbrace{\sum_{q =1}^Q\beta_{n,m,q}\sqrt{\mathrm{PL}(d_{n,m,q})} e^{-j2\pi (i \tau_{n,m,q}\Delta f-t f_{\mathrm{D},n,m,q}\Delta T)}  
 \bfa_{\mathrm{r}}( \vartheta_{m,q})\bfa_{\mathrm{t}}^\mathrm{H}(\theta_{n,q})}_{\overset{\Delta}{=} ~\mathbf{G}_{i,n,m}[t]}\mathbf{x}_{i, n}[t]  + \mathbf{z}_{i,m}[t]. 
\label{eq:echo}
\ee
\end{figure*}

Let $N_\mathrm{s}$ and $\Delta f$ denote the number of subcarriers and the subcarrier spacing, respectively. The OFDM symbol duration is given by $\Delta T = 1/\Delta f + T_{\mathrm{p}}$, where $T_{\mathrm{p}}$ is the duration of the cyclic prefix. 
Let $\mathbf{s}_{i}[t]  = (s_{i,1}[t], \ldots, s_{i, K}[t]) \in \mathbb{C}^{K}$ denote the $t$-th transmit vector for the $K$ users on the $i$-th OFDM subcarrier, where $i = 0, \ldots, N_{\mathrm{s}}-1$, $t = 1, \ldots, T_{\mathrm{s}}$, and $T_{\mathrm{s}}$ denotes the  number of OFDM symbols. 
We assume each element of vector $\mathbf{s}_{i}[t]$ has unit power and the transmit symbols are statistically independent, i.e., $\mathbb{E}\{\mathbf{s}_{i}[t]\mathbf{s}^\mathrm{H}_{i}[t]\} = \mathbf{I}_{K}$. 
Let $\mathbf{x}_{i, n}[t] \in \mathbb{C}^{N_{\mathrm{t}}}$ denote the signal on the $i$-th subcarrier transmitted by the $n$-th transmit AP during the $t$-th OFDM symbol interval.  It can be expressed as 
\be
\mathbf{x}_{i, n}[t] = \sum_{k=1}^K \mathbf{w}_{i,n,k}s_{i,k}[t] = \mathbf{W}_{i, n}\mathbf{s}_{i}[t],  \label{eq:trans sig}
\ee 
where $\mathbf{w}_{i,n,k} \in \mathbb{C}^{N_{\mathrm{t}}}$ is the precoder for the $k$-th user assigned to the $n$-th transmit AP for transmission on the $i$-th subcarrier, and   
$\mathbf{W}_{i, n}\overset{\Delta}{=} [\mathbf{w}_{i,n,1} \cdots \mathbf{w}_{i,n,K}]\in \mathbb{C}^{N_{\mathrm{t}}\times K}$.  
The transmit power of the $n$-th transmit AP is given by $\sum_{i=0}^{N_{\mathrm{s}}-1} \|\mathbf{W}_{n,i}\|^2_F$. Let $P$ denote the maximum power at each transmit AP. We have  $\sum_{i=0}^{N_{\mathrm{s}}-1} \|\mathbf{W}_{n,i}\|^2_F\leq P$. The precoded signals in (\ref{eq:trans sig}) are then transformed into time domain signals via inverse discrete Fourier transform (IDFT) and a cyclic prefix of period $T_{\mathrm{p}}$ is inserted to mitigate inter-symbol interference. The time domain signals are assigned to the corresponding transmit APs. After digital-to-analog conversion and RF conversion, the RF signals are emitted with carrier frequency $f_{\mathrm{c}}$ by the transmit AP antennas.

\subsection{Communication Model}

Let $\hcomm \in \mathbb{C}^{N_{\mathrm{t}}}$ denote the communication channel vector between the $n$-th transmit AP and the $k$-th user on the $i$-th subcarrier.  
We stack the channels between the $k$-th user and all transmit APs on the $i$-th subcarrier as $\hstack  = [(\mathbf{h}_{i,1, k})^\mathrm{T}~ \cdots ~ (\mathbf{h}_{i,N, k})^\mathrm{T}]^\mathrm{T} \in \mathbb{C}^{NN_{\mathrm{t}}}$. 
Similarly, by stacking the beamforming vectors for the $k$-th user on the $i$-th subcarrier of all transmit APs, we obtain the beamforming vector for the $k$-th user on the $i$-th subcarrier as $
\mathbf{w}_{i,k} = [\mathbf{w}^\mathrm{T}_{i,1, k}~ \ldots ~ \mathbf{w}^\mathrm{T}_{i,N, k}]^\mathrm{T} \in \mathbb{C}^{NN_{\mathrm{t}}}$. After down-conversion, analog-to-digital conversion, cyclic prefix  removal, and DFT, 
the received signal at the $k$-th user on the $i$-th subcarrier during the $t$-th OFDM symbol interval can be written as
\bea
&&\hspace{-1.2 cm} y^{(\mathrm{c})}_{i,k} [t] =  \sum_{n=1}^{N} (\hcomm)^\mathrm{H} \mathbf{x}_{i, n}[t] + n_{i,k}[t] \\ &&\hspace{-0.15 cm} = 
\underbrace{(\hstack)^\mathrm{H}\mathbf{w}_{i,k}s_{i,k}[t]}_{\text{Desired signal}} \nonumber \\  &&\hspace{0.4 cm} + \underbrace{\sum^K_{l=1, l\neq k} (\hstack)^\mathrm{H}\mathbf{w}_{i,l}s_{i,l}[t]}_{\text{Combined interference}} + \underbrace{n_{i,k}[t]}_{\text{Noise}},
\eea 
where  $n_{i,k}[t] \sim \mathcal{CN}(0, \sigma_{\mathrm{c}}^2)$ is the received noise of the $k$-th user on the $i$-th subcarrier. 
Conventional MIMO beamforming techniques, such as maximum ratio transmission, zero-forcing, and minimum mean-square error (MMSE) beamforming, can be employed for the design of $\mathbf{w}_{i,k}$. In practice, the channel vector between each transmit AP and each user can be estimated through uplink training. In this work, we assume that the CPU has perfect knowledge of the CSI and employs centralized MMSE beamforming\footnote{Similar to \cite{ZXiao-TSP,ML-TAES20}, we focus on target sensing given a fixed transmit beamforming design. The assumption of perfect CSI and adoption of communication-centric beamformers leads to a communication performance upper bound. Imperfect CSI has minimal impact on sensing performance. Our proposed scheme can also be applied in combination with other transmit beamforming algorithms and under imperfect CSI conditions.} for the transmit APs to effectively mitigate multiuser interference.

\subsection{Sensing Model}

The transmit signals in (\ref{eq:trans sig}) are reflected by the targets within the area of interest and the reflected sensing signals are collected by the receive APs. 
Similar to \cite{ZX-TSP}, \cite{ZXiao-TSP}, and \cite{ML-TAES20}, we assume that there is a line-of-sight (LoS) path between each transmit/receive AP and each target\footnote{We assume the contributions of the multipath components are small. For simplicity, we do not consider their impact on sensing channel modeling. However, we evaluate the impact of multipath components on the sensing performance via simulations in Section \ref{sec:multipath simulation}.}. After sampling and DFT processing,  
the received sensing signal during the $t$-th OFDM symbol interval on the $i$-th subcarrier at the $m$-th receive AP is given by (\ref{eq:echo}) shown at the bottom of this page. 
In (\ref{eq:echo}),   $\beta_{n,m,q}\sim \mathcal{CN}(0, \chi^2)$ is a complex reflection coefficient, which includes the effects due to small-scale pathloss and radar cross section of the $q$-th target \cite{UD-TCOM}. $\mathrm{PL}(d_{n,m,q}) = \alpha_0(d_{n,m,q}/d_0)^{-\zeta}$ is the large-scale LoS pathloss coefficient between the $n$-th transmit AP and the $m$-th receive AP via the $q$-th target,  where  $\alpha_0$ is the pathloss at reference distance $d_0$ and $\zeta$ is the pathloss exponent. 
$d_{n,m,q} = d_{n,q} + d_{m,q}$ is the bistatic range measured from the $n$-th transmit AP, via the $q$-th target, to the $m$-th receive AP, where $d_{n,q}$ and $d_{m,q}$ are given as follows:
\setcounter{equation}{6}
\bea 
 d_{n,q} = \|\mathbf{t}_n - \mathbf{g}_q\|_2, ~~  d_{m,q}  = \|\mathbf{g}_q - \mathbf{r}_m\|_2. \label{eq:dist}
\eea 
$\theta_{n,q}$ corresponds to the AoD of the $q$-th target at the $n$-th transmit AP.  $\vartheta_{m,q}$ denotes the AoA of the $q$-th target at the $m$-th receive AP.  
$\mathbf{z}_{i,m}[t] \sim \mathcal{CN}(0, \xi_{\mathrm{z}}^2\mathbf{I}_{M_{\mathrm{r}}})$ is the observed noise at the $m$-th receive AP on the $i$-th subcarrier during the $t$-th OFDM symbol interval. $\tau_{n,m,q}$ and $f_{\mathrm{D}, n,m,q}$ are the bistatic delay and Doppler frequency shift associated with the $n$-th transmit AP and the $m$-th receive AP via the $q$-th target, respectively. They are defined as follows:
\bea 
& \tau_{n,m,q} = \frac{d_{n,m, q}}{c}, \\
& f_{\mathrm{D}, n,m,q} = \frac{(v_{n,q} \hspace{0.05 cm}+\hspace{0.05 cm} v_{m,q})}{c}f_{\mathrm{c}}, 
\eea 
where $v_{n,q}$ and  $v_{m,q}$ are the radial velocities of the $q$-th target w.r.t. the $n$-th transmit AP and the $m$-th receive AP, respectively. If we consider the deployment shown in Fig. \ref{fig:coordinate}(a), where the transmit ULA and receive ULA are oriented with a $180^{\circ}$ phase shift relative to each other, the radial velocities can be expressed as follows\footnote{When the transmit and receive ULAs are deployed with different orientations, the expressions for the radial velocity change accordingly.}:
\bea 
&v_{n,q} = - v_q^{\mathrm{x}}\cos(\theta_{n,q}) + v_q^{\mathrm{y}} \sin(\theta_{n,q}), \\ 
& v_{m,q} = v_q^{\mathrm{x}}\cos(\vartheta_{m,q}) - v_q^{\mathrm{y}} \sin(\vartheta_{m,q}).
\eea 

We note that the received sensing signal in (\ref{eq:echo}) contains information about angles (via the AoAs), ranges (via the delays), and radial velocities (via the Doppler frequency shifts).
Our goal is to estimate the location and velocity of the targets by leveraging the sensing signals obtained from multiple receive APs. Conventional fully distributed sensing approaches \cite{zhang-TCOM,QS-JSAC,ML-TAES20}  may suffer from performance degradation due to errors in the parameters estimated by each receive AP. On the other hand, DNN-based centralized sensing schemes \cite{ZW-ICASSP} can learn to directly map the reflected sensing signals to the target's location and velocity, offering higher-accuracy estimates even in noisy environments. 
However, this approach incurs a large fronthaul signaling overhead for forwarding the sensing signals from the receive APs to the CPU. To address this issue while guaranteeing high sensing accuracy, in the next section, we propose a D-VQVAE network for collaborative processing by the receive APs and the CPU.

\section{D-VQVAE for Cooperative ISAC-Assisted Localization and Velocity Estimation \label{sec:vqvae}}

To balance the sensing accuracy and fronthaul signaling overhead in cooperative ISAC systems, we propose a collaborative processing scheme for target sensing where the overall task is split between the receive APs and the CPU. Instead of transmitting high-dimensional raw sensing signals to the CPU, each receive AP first performs signal compression, sensing-related feature extraction, and quantization locally. The quantization results are then forwarded to the CPU. The information from all receive APs is fused at the CPU for target localization and velocity estimation. 

To facilitate effective compression while extracting essential information for target sensing, we leverage deep learning techniques. VAEs \cite{vae} are commonly used to reduce the dimensionality of input vectors or tensors by mapping them to continuous latent representations with reduced dimensions. 
VQVAEs \cite{vqvae} incorporate a vector quantization module into the VAE framework, enabling the encoding of inputs into discrete latent vectors suitable for efficient digital transmission. By further extending the VQVAE network,  we propose a D-VQVAE network for collaborative processing by the receive APs and the CPU in an end-to-end manner.
Specifically, each receive AP first encodes the received reflected sensing signals into continuous latent features, which are then quantized into a set of discrete latent vectors using a codebook. The codebook enables efficient compression by representing the continuous features as a reduced set of discrete codewords. Each receive AP forwards only the indices of these codewords to the CPU. Based on the indices obtained from all receive APs, the CPU recovers the discrete latent vectors and uses a decoder to estimate the location and velocity of the targets. 
In the following, we provide the details of the encoder, the codebook-based vector quantization, and the decoder within the D-VQVAE network.

\begin{figure*}[t]
\centering
\includegraphics[width=7.0 in]{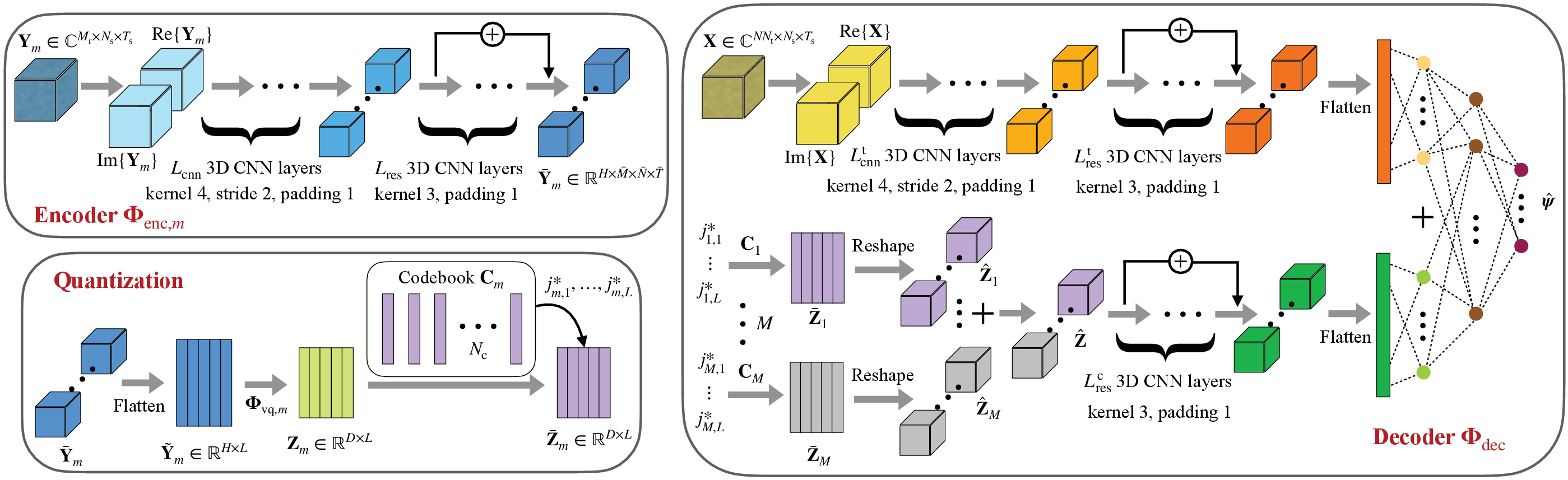}
\caption{The architecture of the proposed D-VQVAE network. Each receive AP uses an encoder to encode the obtained sensing signals locally, followed by vector quantization based on a codebook. The indices of the selected codewords are forwarded to the CPU. The CPU recovers the discrete latent feature vectors based on the indices received. Finally, the location and velocity of the targets are estimated by a decoder.}
\label{fig:vqvae}
\end{figure*} 

\subsection{Encoder Design}

An encoder is deployed at each receive AP for signal compression and feature extraction. 
Given the reflected sensing signals, shown in (\ref{eq:echo}), at each receive AP, we first concatenate these signals across all the $N_{\mathrm{s}}$ subcarriers and obtain matrix $\mathbf{Y}_m[t] = [\mathbf{y}_{0,m}[t] ~ \cdots ~ \mathbf{y}_{N_{\mathrm{s}}-1,m}[t]] \in \mathbb{C}^{M_{\mathrm{r}}\times N_{\mathrm{s}}}$ for the $m$-th receive AP. We further aggregate the sensing signals for all $T_{\mathrm{s}}$ OFDM symbol intervals and denote the resulting three-dimensional (3D) tensor by $\mathbf{Y}_m = [\mathbf{Y}_m[1] ~ \cdots ~ \mathbf{Y}_m[T_{\mathrm{s}}]] \in \mathbb{C}^{M_{\mathrm{r}}\times N_{\mathrm{s}} \times T_{\mathrm{s}}}$. We extract the real and imaginary parts of $\mathbf{Y}_m$, which are given by $\mathrm{Re}\{\mathbf{Y}_m\}$ and $\mathrm{Im}\{\mathbf{Y}_m\}$, respectively.  
Then, $\mathrm{Re}\{\mathbf{Y}_m\}$  and $\mathrm{Im}\{\mathbf{Y}_m\}$ are normalized and used as the input to the encoder of the $m$-th receive AP.

The reflected sensing signals contain information across the space, frequency, and time domains, which is critical for effective feature extraction. To capture these joint 3D domain features, we employ 3D CNNs in the encoder to downsample the 3D signals and extract sensing-related features. 
Specifically, the real and imaginary parts of the concatenated reflected sensing signals, i.e., $\mathrm{Re}\{\mathbf{Y}_m\}$ and $\mathrm{Im}\{\mathbf{Y}_m\}$, are regarded as two input channels for 3D convolution. At the $m$-th receive AP, where $m=1, \ldots, M$, $L_{\mathrm{cnn}}$ 3D CNN layers are employed to downsample the input tensor. The kernel size is set to $4$ with stride $2$ and padding $1$ to achieve a downsampling rate of $2$. The trainable parameters of the CNN layers of the encoder at the $m$-th receive AP are collected in $\bm\Phi_{\mathrm{cnn}, m}$. 
The number of output channels of the last layer is denoted by $H$. After downsampling, we employ a residual network comprising $L_{\mathrm{res}}$ 3D CNN layers with an additional identity mapping to extract the sensing-related features. This residual network enables us to learn deep space-frequency-time domain correlations without suffering from vanishing gradients, allowing the encoder to capture essential sensing-related features. The kernel size is set to $3$ with padding $1$.
We denote the parameters of the residual 3D CNN layers by $\bm \Phi_{\mathrm{res},m}$. We use the rectified linear unit (ReLU) as the activation function for all the CNN layers. 
Then, we obtain the encoded continuous latent features, $\bar{\mathbf{Y}}_m \in \mathbb{R}^{H \times \bar{M} \times \bar{N}\times \bar{T}}$, where $\bar{M}$, $\bar{N}$, and $\bar{T}$ are the reduced dimensions of signals after encoding.
Let $\mathcal{E}_{m}(\cdot; \bm \Phi_{\mathrm{enc},m})$ denote the encoder at the $m$-th receive AP, where $\bm \Phi_{\mathrm{enc},m} = \{\bm\Phi_{\mathrm{cnn},m}, \bm\Phi_{\mathrm{res},m}\}$ collects all parameters. The encoded continuous latent features can be expressed as 
\bea 
\bar{\mathbf{Y}}_m = \mathcal{E}_{m}(\mathbf{Y}_m; \bm \Phi_{\mathrm{enc},m}). \label{eq:encoded feature}
\eea 
The architecture of the proposed encoder is illustrated in the top-left part of Fig. \ref{fig:vqvae}.

\subsection{Codebook-Based Vector Quantization}

Given the continuous latent features in (\ref{eq:encoded feature}), each receive AP then employs a vector quantization scheme to transform the continuous features into discrete latent features. This transformation is achieved by quantizing the encoded continuous features in (\ref{eq:encoded feature}) based on a codebook. Only the indices of the selected codewords are forwarded to the CPU, which can effectively reduce the signaling overhead on the fronthaul link. In particular, let $\mathbf{C}_m = [\mathbf{c}_{m,1} ~ \cdots ~ \mathbf{c}_{m,N_{\mathrm{c}}}]\in \mathbb{R}^{D\times N_{\mathrm{c}}}$ denote the codebook of the $m$-th receive AP for vector quantization, which is shared with the CPU. The codebook consists of $N_{\mathrm{c}}$ codewords, where each codeword $\mathbf{c}_{m,j}$ has $D$ dimensions. Given $\bar{\mathbf{Y}}_m$, which contains encoded features of size $\bar{M} \times \bar{N}\times \bar{T}$ with each feature represented by a continuous vector of size $H$, we aim to quantize each continuous feature vector into a discrete latent vector based on the codebook.

In particular, $\bar{\mathbf{Y}}_m$ is first flattened into $\tilde{\mathbf{Y}}_m\in\mathbb{R}^{H\times  L}$, where $L = \bar{M} \bar{N}\bar{T}$ denotes the number of continuous feature vectors.  
We apply a linear projector with weight $\bm \Phi_{\mathrm{vq},m} \in \mathbb{R}^{D\times  H}$ to transform $\tilde{\mathbf{Y}}_m$ into $\mathbf{Z}_m = [\mathbf{z}_{m,1} ~ \cdots ~ \mathbf{z}_{m,L}] \in \mathbb{R}^{D \times L}$ to match the dimensionality to that of the codebook. This transformation can be expressed as follows:
\bea
\mathbf{Z}_m = \bm \Phi_{\mathrm{vq},m} \tilde{\mathbf{Y}}_m. \label{eq:pre_vq}
\eea 
For each vector $\mathbf{z}_{m,l}$, we use a quantizer $\mathcal{Q}(\cdot; \mathbf{C}_m)$ to compare it with all the codewords in codebook $\mathbf{C}_m$, and choose the one which is nearest to it in terms of the Euclidean distance as the quantization output, $\bar{\mathbf{Z}}_{m} = [\bar{\mathbf{z}}_{m,1}~ \cdots ~ \bar{\mathbf{z}}_{m,L}]$:
\bea
&\hspace{0.4 cm} \bar{\mathbf{Z}}_{m} = \mathcal{Q}(\mathbf{Z}_{m}; \mathbf{C}_m), ~ \mathrm{where}~ \bar{\mathbf{z}}_{m,l} = \mathbf{c}_{j_{m,l}^{\ast}}, \label{eq:quantize} \\ 
& \hspace{-1.3 cm} \mathrm{and} ~ j_{m,l}^{\ast} = \arg \underset{1\leq j\leq N_{\mathrm{c}}}{\min}  \|\mathbf{z}_{m,l} - \mathbf{c}_{m,j}\|_2^2.  \label{eq:index}
\eea 
Then, the $m$-th receive AP sends these indices $j_{m,l}^{\ast}$,  $l=1, \ldots, L$,  back to the CPU. The codebook-based vector quantization process is shown in the bottom-left part of Fig. \ref{fig:vqvae}.

\subsection{Decoder Design}

On the CPU side, after it has obtained the indices from all the receive APs, the discrete latent feature matrix, $\bar{\mathbf{Z}}_m = [\bar{\mathbf{z}}_{m,1} ~ \cdots ~ \bar{\mathbf{z}}_{m,L}] \in \mathbb{R}^{D\times L}$, $m = 1, \ldots, M$, can be reconstructed. The CPU uses a decoder to estimate the location and velocity of the targets based on the discrete latent feature matrix and the transmit signals in (\ref{eq:trans sig}).

\begin{figure}[t]
\centering
\includegraphics[width=2.4 in]{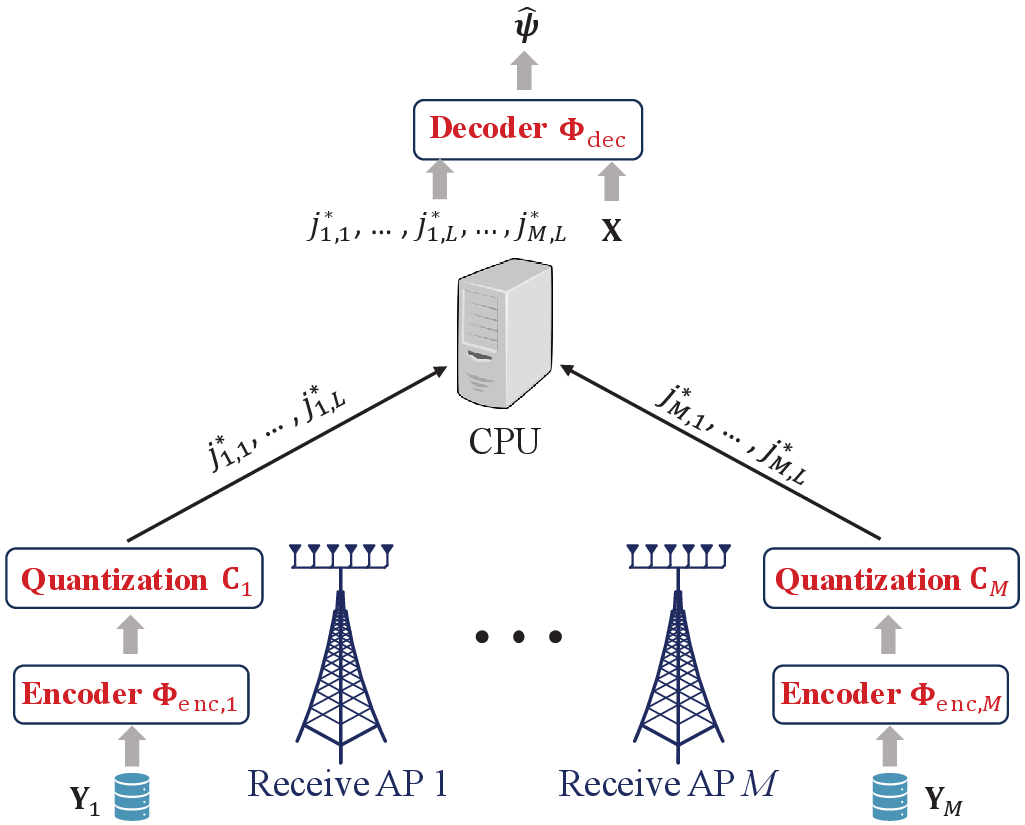}
\caption{Each AP encodes the reflected sensing signals locally through its encoder, then quantizes the encoded feature using a codebook. The CPU estimates the location and velocity of the targets through a decoder.}
\label{fig:flow}
\end{figure}

In particular, we reshape the discrete latent feature matrix, $\bar{\mathbf{Z}}_m$, of the $m$-th receive AP into tensor $\hat{\mathbf{Z}}_m\in \mathbb{R}^{D\times  \bar{M} \times\bar{N}\times\bar{T}}$, which spans the space-frequency-time domain. We further concatenate the feature tensors of all the receive APs along the first dimension and obtain $\hat{\mathbf{Z}} \in \mathbb{R}^{DM \times \bar{M} \times\bar{N}\times\bar{T}}$. We employ a residual network with $L_{\mathrm{res}}^{\mathrm{c}}$ 3D CNN layers for feature processing across the space-frequency-time domain. The corresponding network parameters are collected in matrix $\bm\Phi_{\mathrm{res}}^{\mathrm{c}}$.
Furthermore, given the transmit signals in (\ref{eq:trans sig}), which are available at the CPU, we construct a 3D tensor $\mathbf{X} \in \mathbb{C}^{NN_{\mathrm{t}}\times N_{\mathrm{s}} \times T_{\mathrm{s}}}$, which aggregates the transmit OFDM signals across the $N$ transmit APs, $N_{\mathrm{s}}$ subcarriers, and  $T_{\mathrm{s}}$ OFDM symbol intervals. The real and imaginary parts of $\mathbf{X}$ are denoted as $\mathrm{Re}\{\mathbf{X}\}$ and $\mathrm{Im}\{\mathbf{X}\}$, respectively, which are regarded as two channels for 3D convolution. 
Similar to the processing of the reflected sensing signals, we employ $L^{\mathrm{t}}_{\mathrm{cnn}}$ 3D CNN layers to downsample the concatenated transmit signal $\mathbf{X}$. 
The parameters of the stacked 3D CNN layers are collected in matrix $\bm\Phi_{\mathrm{cnn}}^{\mathrm{t}}$.
After downsampling, we employ a residual network comprising $L^{\mathrm{t}}_{\mathrm{res}}$ 3D CNN layers with an additional identity mapping to further extract the features from the space-frequency-time domain. 
The outputs of the residual networks are flattened. We then apply linear projectors with weight matrices $\bm\Phi^{\mathrm{c}}_{\mathrm{fc}}$ and $\bm\Phi^{\mathrm{t}}_{\mathrm{fc}}$ to the flattened vectors, respectively, to extract the combined and high-level features. The outputs are concatenated and fed into a fully connected layer with weight matrix $\bm\Phi_{\mathrm{fc}}$.
Finally, we employ another fully connected layer with weight matrix $\bm\Phi_{\mathrm{out}}$ to generate the estimated locations $\hat{\mathbf{g}}\in \mathbb{R}^{2Q}$ and velocities $\hat{\mathbf{v}} \in \mathbb{R}^{2Q}$ for all the targets. 
Let $\mathcal{D}(\cdot; \bm\Phi_{\mathrm{dec}})$ denote the decoder at the CPU, where $\bm\Phi_{\mathrm{dec}} = \{\bm\Phi^{\mathrm{c}}_{\mathrm{res}}, \bm\Phi^{\mathrm{t}}_{\mathrm{cnn}}, \bm\Phi^{\mathrm{c}}_{\mathrm{fc}}, \bm\Phi^{\mathrm{t}}_{\mathrm{fc}}, \bm\Phi_{\mathrm{fc}}, \bm\Phi_{\mathrm{out}}\}$ contains the decoder parameters. The estimated results can be expressed as
\bea
\hat{\bm\psi} = \mathcal{D}\left(\{\bar{\mathbf{Z}}_m\}_m, \mathbf{X}; \bm\Phi_{\mathrm{dec}}\right), 
\eea 
where $\hat{\mathbf{g}} = \hat{\bm\psi}[1:2Q]$ and $\hat{\mathbf{v}} = \hat{\bm\psi}[2Q+1:4Q]$ are the estimated location and velocity of the targets. 
For the $q$-th target, the estimated location and velocity are given by $\hat{\mathbf{g}}_q = \hat{\mathbf{g}}[2(q-1)+1: 2q]$ and $\hat{\mathbf{v}}_q = \hat{\mathbf{v}}[2(q-1)+1: 2q]$, respectively. 
The architecture of the decoder is shown in the right part of Fig. \ref{fig:vqvae}. The workflow for the collaborative processing is illustrated in Fig. \ref{fig:flow}.

\subsection{Signaling Overhead Analysis \label{sec:overhead}}

In this subsection, we analyze the fronthaul signaling overhead incurred by our proposed scheme and compare it with the fully distributed and centralized sensing schemes. 

For the fully distributed sensing scheme, each receive AP transmits three parameters, i.e., angle, range, and radial velocity, for each target to the CPU. Suppose each parameter is represented as a single-precision floating-point number, requiring $32$ bits for encoding. With $Q$ targets, the total fronthaul signaling overhead for each receive AP is equal to $96Q$ bits, which scales linearly with the number of targets. If the parameter update frequency is $f$ updates per second, then the transmission rate will be equal to $96Qf$ bits per second (bit/s). However, as we mentioned earlier, this approach may suffer from estimation error propagation and limited sensing accuracy, as each receive AP operates independently without leveraging spatial relationships with other receive APs. 

On the other hand, the centralized sensing approach requires each receive AP to send the raw sensing signals, i.e., $\mathbf{Y}_m$ with dimension $M_{\mathrm{r}} \times N_{\mathrm{s}} \times T_{\mathrm{s}}$, directly to the CPU for joint processing. Note that $\mathbf{Y}_m$ is a complex tensor. Assuming that each element of the real and imaginary parts of $\mathbf{Y}_m$ is represented by a $32$-bit floating-point number, the total fronthaul signaling overhead is given by $64M_{\mathrm{r}}  N_{\mathrm{s}} T_{\mathrm{s}}$ bits. Given a sensing signal update frequency of $f$ updates per second, the required transmission rate is $64M_{\mathrm{r}}  N_{\mathrm{s}} T_{\mathrm{s}}f$ bit/s.  In practice, achieving high sensing accuracy necessitates a large number of antennas, subcarriers, and OFDM symbols, making $\mathbf{Y}_m$ typically very high-dimensional.
This results in substantial fronthaul signaling overhead due to the transmission of $\mathbf{Y}_m$, leading to increased latency and bandwidth consumption. 
Therefore, while centralized sensing can achieve higher sensing accuracy by leveraging global sensing information, the significant overhead limits its scalability and efficiency.

Our proposed scheme provides a trade-off between the fully distributed and centralized sensing approaches. Instead of sending high-dimensional raw sensing signals, each receive AP performs local encoding of the received sensing signals followed by codebook-based quantization. Only the indices of the selected codewords are forwarded to the CPU. This can significantly reduce the amount of information exchanged over the fronthaul links. Specifically, for a codebook of size $N_{\mathrm
{c}}$, $N_{\mathrm{b}} = \log_2 N_{\mathrm{c}}$ bits are required to index each codeword. The fronthaul signaling overhead incurred by each receive AP is equal to $N_{\mathrm{b}}L$ bits, where $L$ is the number of discrete feature vectors, which depends on the hyperparameters of the D-VQVAE network. Similar to the fully distributed and centralized sensing approaches, given the update frequency of $f$ updates per second, the required transmission rate on the fronthaul link is equal to $N_{\mathrm{b}}Lf$ bit/s. 
We note that the values of $N_{\mathrm{b}}$ and $L$  are significantly smaller than the original dimensions of the sensing signals. 
Hence, the signaling overhead compared with the centralized approach is significantly reduced while the useful sensing information necessary for accurate target localization and velocity estimation is preserved. 

\section{Collaborative Learning-Assisted Training of the D-VQVAE Network \label{sec:learning}}

To facilitate offline training of the proposed D-VQVAE network, we apply a collaborative learning framework to jointly train the receive AP-side networks (i.e., encoder and codebook) and the  CPU-side network (i.e., decoder). The offline training consists of a forward propagation phase followed by backpropagation. The receive APs first encode their respective sensing signals locally. Then, after vector quantization, each receive AP forwards the indices to the CPU. The CPU reconstructs the discrete latent feature vectors based on the indices and estimates the location and velocity of the targets through the decoder. This completes the forward propagation process. For backpropagation, we note that the quantization operation is non-differentiable, preventing the gradients from propagating directly through the discrete codebook-based quantization step. To address this issue,  we apply the straight-through estimator \cite{vqvae}, which allows the gradients to bypass the non-differentiable step. In particular, the CPU calculates the loss gradients w.r.t. the decoder’s inputs and directly propagates these gradients to the encoders at the distributed receive APs.

For offline training, we construct a training dataset $\mathcal{D}$, which contains $N_{\mathrm{d}}$ data samples. Each data sample consists of a pair of inputs and labels. The transmit OFDM signals $\mathbf{X}$ and the reflected sensing signals $\mathbf{Y}_m$,  $m=1, \ldots, M$, serve as input, while the true location and velocity of the targets $\bm \psi$ are the labels. We denote the training dataset as $\mathcal{D} = \{\mathbf{X}^{(d)}, \mathbf{Y}_1^{(d)}, \ldots, \mathbf{Y}_M^{(d)}, \bm\psi^{(d)}\}_{d=1}^{N_{\mathrm{d}}}$.  
The $m$-th receive AP only has knowledge of the reflected sensing signals that it has received, i.e., $\mathbf{Y}_m$, $m=1, \ldots, M$. The transmit OFDM signal $\mathbf{X}$ is available at the CPU. During training, we assume that the true location and velocity of the targets, $\bm\psi$, are available at the CPU. We further assume that the index of the data samples in $\mathcal{D}$ is known by both the receive APs and the CPU, and the sampling mechanism is predefined by the CPU and shared with all the receive APs. Therefore, the network inputs are paired with the labels during training. 
The proposed D-VQVAE network is trained using the Adam optimizer \cite{Adam}. During the offline training phase, the labels $\{\bm\psi^{(d)}\}_{d=1}^{N_{\mathrm{d}}}$ are normalized using max-min normalization. 
The CPU records the maximum and minimum values for this process. In the online operation phase, the estimated results are rescaled back to their normal values based on the recorded maximum and minimum values.

\subsection{Update of the Encoders, Codebooks, and Decoder}

The developed D-VQVAE network is collaboratively trained between the CPU and the receive APs, which enables joint optimization of encoders, codebooks, and the decoder.

In particular, we employ the estimation loss $\mathcal{L}_{\mathrm{est}}$ to update the distributed encoders and the decoder by minimizing the discrepancy between the ground truth and the estimated values.  Considering the MSE between them as the estimation loss, the estimation loss can be calculated as follows:
\bea 
\mathcal{L}_{\mathrm{est}} = \sum_{r=1}^{R}\|\bm\Psi^{(r)} - \hat{\bm\Psi}^{(r)}\|_2^2, \label{eq:est loss}
\eea 
where $R$ denotes the total number of training steps. $\bm\Psi^{(r)} = \{\bm\psi^{(r_1)}, \ldots, \bm\psi^{(r_B)}\}$ and $\hat{\bm\Psi}^{(r)} = \{\hat{\bm\psi}^{(r_1)}, \ldots, \hat{\bm\psi}^{(r_B)}\}$ are batches of labels and outputs in the $r$-th training step, respectively, with $B$ being the batch size.

The codebooks are updated via the EMA scheme \cite{vqvae}. Denote $\mathbf{c}^{(r)}_{m,j}$ as the $j$-th codeword of the $m$-th receive AP in the $r$-th training step. Let $\{\mathbf{z}^{(r)}_{m, N_{j_1}}, \ldots, \mathbf{z}^{(r)}_{m, N_{j_r}}\}$ denote the set of continuous feature vectors that have been assigned to $\mathbf{c}^{(r-1)}_{m,j}$, which is the $j$-th codeword in the $(r-1)$-th training step. Note that $N_{j_r}$ is the total number of feature vectors assigned to the $j$-th codeword. Then, the EMA accumulators can be obtained as follows:
\bea 
&& L^{(r)}_{m,j} = \gamma L^{(r-1)}_{m,j} + (1-\gamma) N_{j_r}, \label{eq:ema accum 1}\\
&& \tilde{\mathbf{c}}_{m,j}^{(r)} = \gamma \tilde{\mathbf{c}}_{m,j}^{(r-1)} + (1-\gamma) \sum_{l= N_{j_1}}^{N_{j_r}}\mathbf{z}^{(r)}_{m,l}, \label{eq:ema accum 2}
\eea 
where $L^{(0)}_{m,j}$ is initialized as zero for $m=1, \ldots, M$ and $j=1, \ldots, N_{\mathrm{c}}$. Similarly, $\tilde{\mathbf{c}}_{m,j}^{(0)} $ is initialized as a zero vector. The $j$-th codeword at the $m$-th AP is updated as:
\bea 
\mathbf{c}^{(r)}_{m,j} = \frac{\tilde{\mathbf{c}}_{m,j}^{(r)}}{L^{(r)}_{m,j}}.  \label{eq:update book}
\eea

Moreover, note that the quantization process introduces a mismatch between the output of the encoder, $\mathbf{z}_{m,l}$, and the codeword. This mismatch can make it difficult for the encoder to learn stable representations. To address this issue, we apply a commitment loss $\mathcal{L}_{\mathrm{com}}$ to encourage the encoder output to stay close to the chosen codeword by penalizing deviations: 
\bea 
\mathcal{L}_{\mathrm{com}} = \omega \sum_{r=1}^R\sum_{l=1}^{L}\sum_{m=1}^{M}\left\|\mathbf{z}^{(r)}_{m,l} - \mathrm{sg}[\mathbf{c}^{(r)}_{j_{m,l}^{\ast}}]\right\|_2^2, \label{eq:commit loss}
\eea 
where $\omega$ is a hyperparameter that scales the commitment loss. 
In (\ref{eq:commit loss}), $\mathrm{sg[\cdot]}$ denotes the stop-gradient operator \cite{vqvae}, which is defined as identity during forward computation and has zero partial derivatives. This can effectively restrict its operand to remain constant without update during backpropagation. 
The stop-gradient operator is applied to each codeword such that only the output of the encoder is being updated. This loss term ensures that the encoder commits to producing outputs that are close to the discrete codebook and helps the encoder learn to map the input consistently to the learned discrete latent space. 

\vspace{0. in}

\subsection{Training Procedure}

In the following, we explain the collaborative learning-assisted offline training step by step. The overall training procedure is summarized in Algorithm \ref{alg:training}.

\begin{algorithm}[t]
\small{
\caption{Collaborative Learning-Assisted Training}
\label{alg:training}
\begin{algorithmic}[1]
\STATE \textbf{Input:} Training dataset $\mathcal{D}$, learning rate of the Adam optimizer, batch size $B$, and total number of training epochs $E$.
\STATE Initialization. 
\FOR{training epoch $e = 1, \ldots,  E$}
\FOR{training step $r = 1, \ldots,  R$}
\STATE Sample a batch of data samples from $\mathcal{D}$. 
\FOR{each receive AP \textbf{in parallel}}
\STATE Local encoding of the reflected sensing signals. 
\STATE Continuous feature quantization using the codebook. 
\STATE Send the discrete latent features (\ref{eq:quantize}) to the CPU. 
\STATE Update its codebook locally. 
\ENDFOR
\STATE  Decoding at the CPU.  
\STATE Decoder update at the CPU. 
\STATE The CPU sends the gradient of the discrete latent features to the corresponding receive APs. 
\FOR{each receive AP \textbf{in parallel}}
\STATE Update its encoder locally. 
\ENDFOR
\ENDFOR
\ENDFOR
\STATE \textbf{Output:} Trained encoders, codebooks, and the decoder. 
\end{algorithmic} } 
\end{algorithm}

\textit{Step 1 - Initialization (line 2)}. The CPU first initializes the parameters for the encoder, codebook, and decoder as $\{\bm\Phi_{\mathrm{enc}, m}^{(0)}\}_{m=1}^M$, $\{\mathbf{C}_{m}^{(0)}\}_{m=1}^M$, and $\bm\Phi_{\mathrm{dec}}^{(0)}$, respectively. Then, the CPU sends the initialized parameters for the encoders and codebooks to the corresponding receive APs. 

\textit{Step 2 - Forward propagation of the receive AP-side network  (lines 7$-$10)}. In the $r$-th training step, each receive AP draws a batch of input data, $\mathcal{Y}_m^{(r)}= \{\mathbf{Y}_m^{(r_1)}, \ldots, \mathbf{Y}_m^{(r_B)}\}$, $m=1, \ldots, M$. The $m$-th receive AP encodes the obtained sensing signals locally and the encoded continuous features are given by $\bar{\mathcal{Y}}_m^{(r)} = \{\bar{\mathbf{Y}}_m^{(r_1)}, \ldots, \bar{\mathbf{Y}}_m^{(r_B)}\}$, which are further transformed and then quantized into discrete latent features $\bar{\mathcal{Z}}_m^{(r)} = \{\bar{\mathbf{Z}}_m^{(r_1)}, \ldots, \bar{\mathbf{Z}}_m^{(r_B)}\}$ based on (\ref{eq:pre_vq}) and (\ref{eq:quantize}), respectively.  Then, each AP forwards the discrete latent features to the CPU\footnote{During training, the codebook at each receive AP is updated locally and is not shared with the CPU in every epoch. Therefore, the discrete latent features have to be sent to the CPU for decoding. During online execution, each receive AP can simply transmit the indices in (\ref{eq:index}) to the CPU.} and update its codebook locally based on (\ref{eq:ema accum 1})$-$(\ref{eq:update book}). 

\textit{Step 3 - Forward propagation of the CPU-side network (line 12)}. The CPU draws a batch of its input data, $\mathcal{X}^{(r)} = \{\mathbf{X}^{(r_1)}, \ldots, \mathbf{X}^{(r_B)}\}$, and performs decoding based on the discrete features obtained from the receive APs and the sampled data $\mathcal{X}^{(r)}$. The batch of the decoder output is given by $\hat{\bm\Psi}^{(r)}= \{\hat{\bm\psi}^{(r_1)}, \ldots, \hat{\bm\psi}^{(r_B)}\}$, where each element represents the estimated locations and velocities given the $r_b$-th input in the batch, $r_b \in \{ r_1, \ldots, r_B\}$. 

\textit{Step 4 - Backpropagation of the CPU-side network (lines 13, 14)}. Next, we update the network parameters by minimizing the estimation loss $\mathcal{L}_{\mathrm{est}}$ in (\ref{eq:est loss}) and the commitment loss $\mathcal{L}_{\mathrm{com}}$ in (\ref{eq:commit loss}) during backpropagation. Given the estimated results $\hat{\bm\Psi}^{(r)}$ and the sampled labels $\bm\Psi^{(r)} = \{\bm\psi^{(r_1)}, \ldots, \bm\psi^{(r_B)}\}$, the gradients of all the layers can be obtained by backpropagation using the chain rule. 
The CPU updates the parameters of the decoder based on the Adam optimizer \cite{Adam} and obtains $\bm\Phi_{\mathrm{dec}}^{(r)}$. When the gradient calculation proceeds to the first layer of the decoder, the CPU sends the gradients of the discrete latent features (i.e., inputs to the decoder) back to the corresponding receive APs. 

\textit{Step 5 - Backpropagation of the receive AP-side network (line 16)}. With the received gradient of the discrete latent features, each receive AP copies the obtained gradients to its encoder output (i.e., continuous latent features). The encoder is then updated through the Adam optimizer by each receive AP locally\footnote{We assume each receive AP is equipped with a graphics processing unit (GPU) to enable efficient local model updates and online inference, and has sufficient memory to store both the encoder network and the codebook.}.

Steps 2$-$5 are iterated for $R$ training steps, and this completes one training epoch. Let $E$ denote the total training epochs. 
After training, we can obtain the trained D-VQVAE network with its optimized decoder parameters $\bm\Phi^{\star}_{\mathrm{dec}}$ for the CPU as well as encoder parameters $\bm\Phi^{\star}_{\mathrm{enc},m}$ and codebook $\mathbf{C}^{\star}_m$ for the $m$-th receive AP, $m=1, \ldots, M$. Each receive AP then sends its trained codebook to the CPU for online execution. 
During online operation, given the reflected sensing signals in (\ref{eq:echo}), each receive AP first encodes the signals locally. After quantizing the encoded features, each receive AP forwards the corresponding indices to the CPU. The CPU reconstructs the discrete latent features and estimates the location and velocity of the targets through the trained decoder. 
The online execution of the target sensing is summarized in Algorithm \ref{alg:testing}.

\begin{algorithm}[t]
\small{\caption{Online Execution for Target Sensing}
\label{alg:testing}
\begin{algorithmic}[1]
\STATE Given $\mathbf{Y}_m$, $m=1, \ldots, M$:
\FOR{each receive AP \textbf{in parallel}}
\STATE Encode the reflected sensing signals and obtain the encoded continuous features as in (\ref{eq:encoded feature}). 
\STATE Reshape the continuous features and apply a linear transformation as in (\ref{eq:pre_vq}). 
\STATE Quantize the features based on the codebook as in (\ref{eq:quantize}). 
\STATE Forward the indices (\ref{eq:index}) to the CPU. 
\ENDFOR
\STATE  The CPU estimates the location and velocity of the targets through the decoder. 
\STATE \textbf{Output:} Estimated location and velocity of the targets $\hat{\bm{\psi}}$.
\end{algorithmic}}  
\end{algorithm} 

\section{Performance Evaluation \label{sec:performance}}

\begin{figure}[t]
    \centering
\includegraphics[width=1.8in]{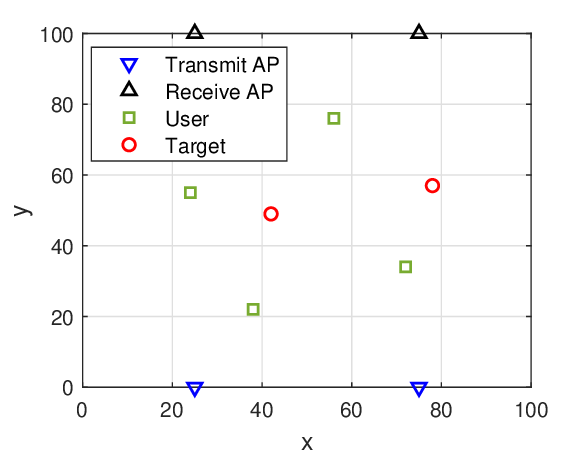}

        \caption{The topology considered in simulations. For different channel realizations, the locations of the APs are fixed, while the users and targets are randomly distributed in the $100\times 100$ m$^2$ area.}
        \label{fig:setting}
        \vspace{-0. cm}
\end{figure}

In this section, we evaluate the sensing performance of the proposed D-VQVAE network through simulations. We consider a cell-free MIMO system with $N=2$ transmit APs and $M=2$ receive APs within a coverage area of $100 \times 100$ m$^2$. Consider a 2D $(\mathrm{x},\mathrm{y})$ coordinate system\footnote{In this work, following existing works \cite{MAI-ICC,ZX-TSP,ZXiao-TSP,ML-TAES20}, we assume all APs and targets lie in a common horizontal plane and the APs employ ULAs without elevation diversity. However, the proposed D-VQVAE network can be extended to 3D scenarios by expanding the network outputs to include the location and velocity coordinates along the $\mathrm{z}$-axis.},  the transmit APs are located at $(25,0)$ and $(75,0)$. The receive APs are placed at $(25,100)$ and $(75,100)$. 
Unless otherwise specified, we consider each AP has $16$ antennas, i.e., $N_{\mathrm{t}} = M_{\mathrm{r}} = 16$. The antenna spacing of the ULAs is set to $d_{\mathrm{t}} = d_{\mathrm{r}} = \lambda_{\mathrm{c}}/2$. 
The transmit APs send OFDM symbols to $K=4$ users for communication. The communication symbols are independently drawn from a $16$-quadrature amplitude modulation ($16$-QAM) constellation. A centralized MMSE beamformer is utilized to precode the transmit symbols. 
We consider there are $Q=2$ targets within this area. The users and targets are assumed to be randomly distributed in the area. The system topology is illustrated in Fig. \ref{fig:setting}. The velocities of each target in $\mathrm{x}$ and $\mathrm{y}$ direction, i.e., $v_q^{\mathrm{x}}$ and $v_q^{\mathrm{y}}$,  are assumed to be between $-20$ m/s and $20$ m/s. 
We list the parameters of the OFDM waveform and the channel conditions in Table \ref{tb:setting}.
Based on this system setting, we generate 20,000 data samples with different channel realizations, where 16,000 of them are used for offline training of the D-VQVAE network, and the remaining 4,000 data samples are used for online testing. The learning rate during training is set to $10^{-4}$. Note that the data samples are normalized during training. During online execution, the estimated results are rescaled back to their normal values.  The hyperparameters are summarized in Table \ref{tb:net}. Each receive AP side encoder has approximately $1.2$ million parameters, and the combined encoder plus codebook occupies about $5$ MB of memory. On the CPU side, the decoder consists of $206$ million parameters, corresponding to a model size of approximately $820$ MB.

\begin{table}[t] 
\centering
\scriptsize{
\caption{System Settings}
{\renewcommand{\arraystretch}{1.0}
\begin{tabular}{|c|c|c|}
\hline
Parameter                           & Symbol & Value                     \\ \hline
Carrier frequency                   &    $f_{\mathrm{c}}$    & $30$ GHz                    \\ \hline
Subcarrier spacing                  &       $\Delta f$ & $240$ kHz                   \\ \hline
Number of subcarriers               &   $N_{\mathrm{s}}$     & $256$                       \\ \hline
Number of OFDM symbols              &  $T_{\mathrm{s}}$      & $256$                       \\ \hline
Cyclic prefix duration              &   $T_{\mathrm{p}}$     & $1.04$  $\mu$s \\ \hline
Reference distance                  &     $d_0$   & $1$ m                       \\ \hline
Pathloss at the reference distance  &   $\alpha_0$     & $-60$ dB                    \\ \hline
Pathloss exponent                   &      $\zeta$  & $2$                       \\ \hline     Variance of the reflection coefficient                   &      $\chi^2$  & $1$                       \\ \hline 
Variance of the sensing noise & $\xi_{\mathrm{z}}^2$ & $-90$ dBm \\ \hline 
\end{tabular}\label{tb:setting}}}
\end{table}

\begin{table}[t] 
\centering
\scriptsize{
\caption{Hyperparameters of D-VQVAE}
{\renewcommand{\arraystretch}{1.0}
\begin{tabular}{|c|c|c|}
\hline
& Hyperparameters & Value \\ \hline
\multirow{3}{*}{Encoder}  & $L_{\mathrm{cnn}}$      &   $2$    \\ \cline{2-3} 
& $L_{\mathrm{res}}$      &     $2$  \\ \cline{2-3} 
& $H$               &     $128$  \\ \hline
\multirow{2}{*}{Codebook} & $D$    &    $512$   \\ \cline{2-3} 
& $N_{\mathrm{c}}$     &    $32$   \\ \hline
\multirow{3}{*}{Decoder}  & $L_{\mathrm{cnn}}^{\mathrm{t}}$    &  $2$     \\ \cline{2-3} 
 & $L_{\mathrm{res}}^{\mathrm{t}}$    &     $2$     \\ \cline{2-3} 
& $L_{\mathrm{res}}^{\mathrm{c}}$      &    $1$   \\ \hline
\end{tabular}\label{tb:net}}}
\end{table}

\vspace{-0.1 in}
\subsection{Baselines and Benchmark} 

For performance comparison, we consider the following five baseline schemes:

\setlength{\hangindent}{2.0em}
$\bullet$ Monostatic sensing by a single BS \cite{ZXiao-TSP}: This scheme assumes that the BS operates in full-duplex mode with perfect self-interference cancellation. The BS transmits OFDM symbols to $K$ users and receives sensing signals reflected by the targets. The BS processes the reflected signals and estimates the angle, range, and radial velocity of the potential targets. For this scheme, we assume the BS is located at $(50, 0)$. 

\setlength{\hangindent}{2.0em}
$\bullet$ CS-based fully distributed sensing scheme \cite{ML-TAES20}: Given the sensing signals collected by the receive APs in (\ref{eq:echo}), this scheme first applies a one-dimensional (1D) CS algorithm to extract the delay information for range estimation. Then, the angle and radial velocity of each target are estimated by each receive AP. The estimated sensing parameters are then sent to the CPU, based on which the location and velocity of each target are obtained.

\setlength{\hangindent}{2.0em}
$\bullet$ MUSIC-based distributed sensing: This scheme extends the monostatic sensing algorithm in \cite{MAI-ICC} to the fully distributed sensing case. Each AP first estimates the angles of the targets via the MUSIC algorithm. Then, range and velocity are estimated through the 2D-DFT estimation method. The location and velocity of each target are determined by collecting the estimated sensing parameters.

\setlength{\hangindent}{2.0em}
$\bullet$ CNN-based centralized sensing scheme \cite{ZW-ICASSP}: In this scheme, the CPU directly estimates the location
and velocity of the targets based on the reflected sensing signals obtained from the receive APs. A DNN architecture is developed, which consists of two 3D CNN layers with a kernel size $5$ for feature extraction, followed by max pooling. The outputs are then flattened and passed through three fully connected layers, each containing $256$ neurons, to generate the final estimates.

\setlength{\hangindent}{2.0em}
$\bullet$ Distributed VAE (D-VAE)-based scheme: This scheme is developed for an ablation study, which does not include the vector quantization module of the D-VQVAE network in Fig. \ref{fig:vqvae}. The encoder and decoder of the D-VAE network retain the same architecture as their counterparts in the D-VQVAE network. The receive APs encode their respective sensing signals locally. Then, the receive APs send the corresponding encoded continuous latent features to the CPU, based on which the location and velocity of each target are estimated at the CPU.

We further derive the CRLB for estimation of $\bm\psi = [\mathbf{g}^{\mathrm{T}} ~ \mathbf{v}^{\mathrm{T}}]^{\mathrm{T}}\in \mathbb{R}^{4Q}$ to serve as a performance benchmark. We denote $\bm\beta = \{\beta_{n,m,q}\}_{n,m,q}$,  which collects the unknown complex coefficients, and $\bm \eta = [\bm\psi^{\mathrm{T}}  ~ \mathrm{Re}\{\bm\beta\}^{\mathrm{T}} ~ \mathrm{Im}\{\bm\beta\}^{\mathrm{T}}]^{\mathrm{T}}\in\mathbb{R}^{4Q+2NMQ}$. 
Given the received sensing signals in (\ref{eq:echo}), we define  the noiseless received sensing signal as
\be\boldsymbol\rho_{i,m}[t]
= \sum^N_{n=1} \mathbf{G}_{i,n,m}[t]\mathbf{x}_{i,n}[t] \in \mathbb{C}^{M_{\mathrm{r}}}. \ee 
Let $\mathbf{F}\in \mathbb{R}^{(4Q+2NMQ)\times (4Q+2NMQ)}$ denote the Fisher information matrix (FIM). 
The element in the $a$-th row and the $b$-th column of $\mathbf{F}$, i.e.,  $[\mathbf{F}]_{a,b}$, is given by \cite{estimation}
\begin{equation}
[\mathbf{F}]_{a,b}
= \mathrm{Re}\left\{\frac{2}{\xi_{\mathrm{z}}^2}\sum_{m=1}^M
\sum_{i = 0}^{N_{\mathrm{s}}-1}\sum_{t=1}^{T_{\mathrm{s}}}
\frac{\partial\boldsymbol\rho_{i,m}[t]^{\mathrm{H}}}{\partial\eta_a}
\frac{\partial\boldsymbol\rho_{i,m}[t]}{\partial\eta_b}\right\},\label{eq:fim}
\end{equation}
where $\eta_a$ $(\eta_b)$ denotes the $a$-th ($b$-th) entry of $\bm\eta$, $a, b = 1, \ldots, 4Q+2NMQ$. The FIM can be represented as 
\[
\mathbf F =
\begin{bmatrix}
\mathbf F_{\psi\psi} & \mathbf F_{\psi\beta}\\
\mathbf F_{\beta\psi} & \mathbf F_{\beta\beta}
\end{bmatrix}, 
\]
with \(\mathbf F_{\beta\psi}=\mathbf F_{\psi\beta}^{\mathrm{T}}\). The partial derivatives of $\bm\rho_{i,m}[t]$ w.r.t. each entry of $\bm\psi$ and $\bm\beta$ are given in Appendix \ref{sec:appendix1}. 
Then, the CRLB matrix for \(\bm\psi\) can be obtained based on the Schur complement as follows:
\be
\mathrm{CRLB}_{\psi}
=\bigl(\mathbf F_{\psi\psi}
-\mathbf F_{\psi\beta}\,\mathbf F_{\beta\beta}^{-1}\,\mathbf F_{\beta\psi}\bigr)^{-1}.
\ee
For the \(q\)-th target, the MSE lower bounds are given as:
\begin{align}
\mathbb E\bigl[(\hat g_q^{x}-g_q^{x})^{2}\bigr]
&\ge
\bigl[\mathrm{CRLB}_{\psi}\bigr]_{2q-1,2q-1},\\
\mathbb E\bigl[(\hat g_q^{y}-g_q^{y})^{2}\bigr]
&\ge
\bigl[\mathrm{CRLB}_{\psi}\bigr]_{2q,2q},\\
\mathbb E\bigl[(\hat v_q^{x}-v_q^{x})^{2}\bigr]
&\ge
\bigl[\mathrm{CRLB}_{\psi}\bigr]_{2Q+2q-1,\,2Q+2q-1},\\
\mathbb E\bigl[(\hat v_q^{y}-v_q^{y})^{2}\bigr]
&\ge
\bigl[\mathrm{CRLB}_{\psi}\bigr]_{2Q+2q,\,2Q+2q}.
\end{align}

\vspace{-0.2 in}
\subsection{Sensing Performance}

\begin{figure}[t]
    \centering
    \subfigure[]{\includegraphics[width=1.65in]{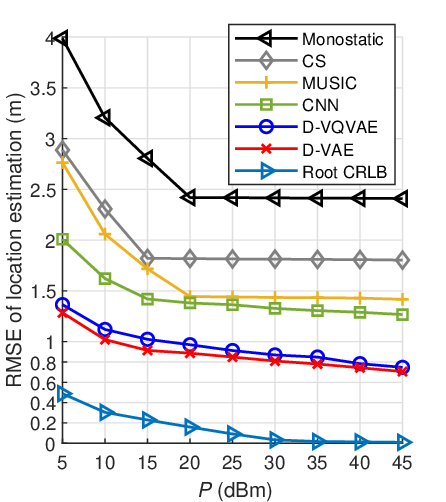}}\hspace{0. in}
    \subfigure[]{
\includegraphics[width=1.65in]{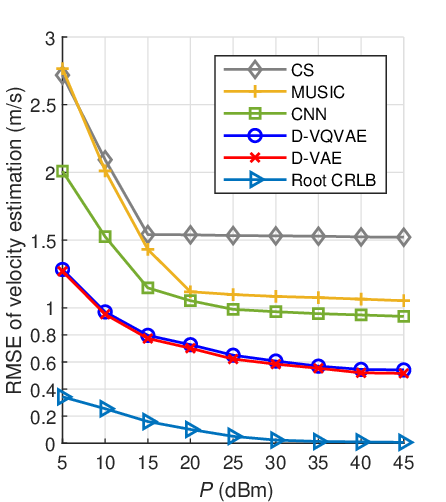}}

        \caption{RMSE for (a) location estimation and (b) velocity estimation versus the maximum transmit power $P$.}
        \label{fig:mse_vs_P}
        \vspace{-0.1 in}
\end{figure}

We evaluate the sensing performance of the considered schemes, where the root mean squared error (RMSE) results for location and velocity estimation and root CRLB are shown in Fig. \ref{fig:mse_vs_P}. For the location estimation results shown in Fig. \ref{fig:mse_vs_P}(a), we include all baseline schemes for performance comparison. In Fig. \ref{fig:mse_vs_P}(b), the performance achieved by the monostatic sensing scheme is not included since only the radial velocity can be estimated by a single BS. The complete velocity vector cannot be obtained. 
In Fig. \ref{fig:mse_vs_P}, the root CRLB shows that the theoretical limits for location and velocity estimation for $P=30$ dBm are under $0.1$ m and $0.1$ m/s, respectively. It can be observed that the RMSE for all schemes decreases with increasing transmit power and eventually saturates to an error floor. This occurs because the available bandwidth and number of OFDM symbols limit the sensing resolution and achievable performance, which causes the RMSE to saturate at a certain level rather than continuing to decrease with further increases in transmit power.  
The results also show that cooperative ISAC yields significant sensing performance improvement compared to monostatic sensing by a single BS. This is because the monostatic sensing scheme relies on a series of DFT operations and point-wise divisions, making it susceptible to noise and less robust in noisy environments. In cooperative ISAC, signals collected from distributed receive APs provide multiview information, which is more reliable than single-point observations. 
When comparing DNN-based schemes (i.e., CNN, D-VAE, and D-VQVAE) with the fully distributed sensing schemes (i.e., CS and MUSIC), the following key advantages of deep learning techniques can be identified. DNN-based schemes leverage the ability to jointly extract features from the sensing signals collected by multiple distributed receive APs. This joint feature extraction enables the network to capture complex space-frequency-time domain patterns, leading to more accurate and robust target localization and velocity estimation. However, for the fully distributed schemes, each receive AP processes the sensing signals independently, without exploiting the spatial dependencies with other receive APs. Moreover, each receive AP estimates the sensing parameters (i.e., angle, range, radial velocity) first before obtaining the location and velocity information of the targets. 
By using DNN, the intermediate parameter estimation stage can be bypassed and the DNNs can determine the location and velocity of the targets directly based on the received sensing signals. Thus, any errors associated with the intermediate stage can be avoided.
When comparing the sensing performance among the three DNN-based schemes, we observe that the CNN-based approach yields higher estimation errors than the other two, despite being a centralized sensing scheme. This limitation stems from the CNN’s inability to fully capture space-frequency-time features, whereas the more advanced DNN architectures can better handle the complex 3D feature extraction. We also notice that the proposed D-VQVAE network provides comparable sensing performance to the D-VAE network without quantization. Although quantization introduces distortion in the latent representations, the collaboratively learned codebooks across receive APs still capture the essential sensing features, and the discrete codebooks yield an implicit quantization regularization effect. Consequently, despite the small quantization error, the developed D-VQVAE network achieves similar RMSE performance as the D-VAE network for both target location and velocity estimation, while significantly reducing the fronthaul signaling overhead.

\begin{figure}[t]
    \centering
    \subfigure[]{\includegraphics[width=1.65in]{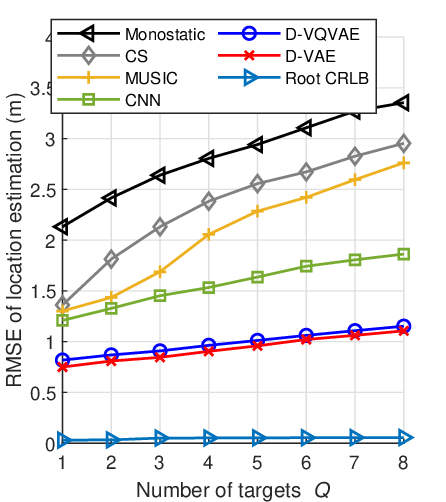}}
    \subfigure[]{
\includegraphics[width=1.65in]{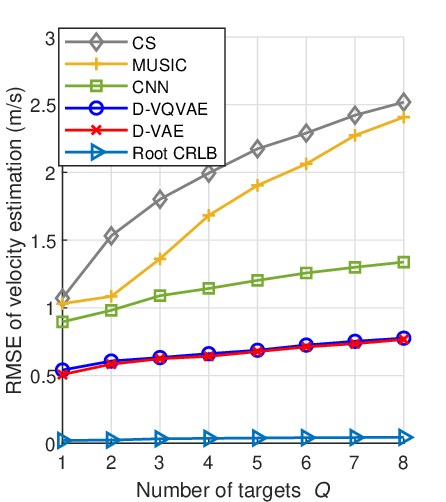}}

        \caption{RMSE for (a) location estimation and (b) velocity estimation versus the number of targets $Q$.}
        \label{fig:mse_vs_Q}
        \vspace{-0.1 cm}
\end{figure}

In Fig. \ref{fig:mse_vs_Q}, we illustrate the dependence of the sensing performance on the number of targets $Q$, with the transmit power $P$ fixed at $30$ dBm. Specifically, we plot the RMSE for both location and velocity estimation. It can be observed that, as the number of targets increases, the conventional approaches (i.e., monostatic sensing and fully distributed sensing schemes) experience significant performance degradation. This is because a larger number of targets introduces higher complexity in separating individual sensing signals within the space-frequency-time domain, leading to higher mutual interference. On the other hand, the DNN-based schemes continue to exhibit satisfactory sensing performance when the number of targets increases. This is because the CPU jointly processes the sensing signals or sensing-related features from all receive APs, leveraging spatial dependencies across distributed APs to enhance target sensing accuracy.
Moreover, the DNN-based schemes are able to learn the complex latent features from both the transmit signals and the reflected sensing signals. By capturing the underlying patterns and correlations within these signals, the DNN architectures can effectively distinguish the sensing signals of individual targets, even in the presence of noise or significant signal overlap. As a result, they maintain robust accuracy and are better suited to handle the increased complexity introduced by a larger number of targets.

\begin{figure}[t]
    \centering
    \subfigure[]{\includegraphics[width=1.65in]{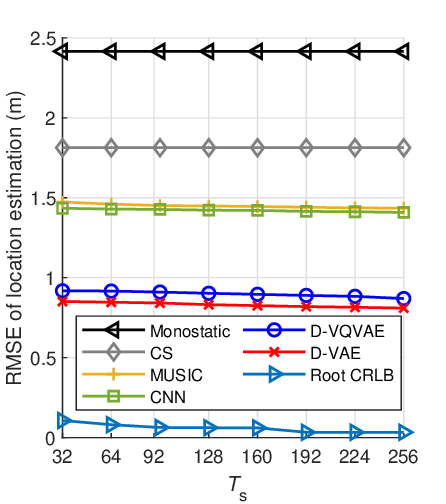}}
    \subfigure[]{
\includegraphics[width=1.65in]{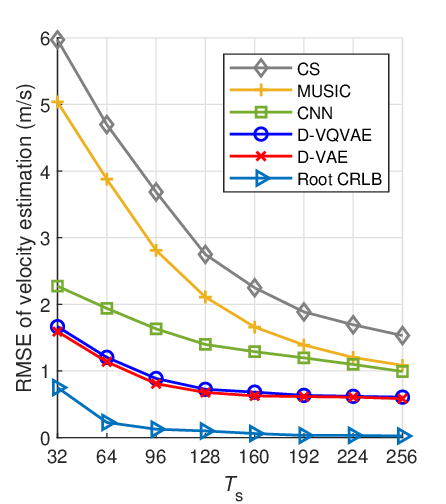}}

        \caption{RMSE for (a) location estimation and (b) velocity estimation versus the number of OFDM symbols $T_{\mathrm{s}}$.}
        \label{fig:mse_vs_Ts}
        \vspace{-0.1 cm}
\end{figure}

In Fig. \ref{fig:mse_vs_Ts}, we investigate how the number of OFDM symbols, $T_{\mathrm{s}}$, affects the sensing performance. We can observe from Fig. \ref{fig:mse_vs_Ts}(a) that varying the value of $T_{\mathrm{s}}$ does not significantly impact the location estimation performance. This is because location estimation relies on the spatial correlations provided by the sensing signals, which can be captured even with a smaller number of OFDM symbols. Increasing $T_{\mathrm{s}}$ mainly enhances the resolution for Doppler estimation, which is critical for velocity estimation. As can be observed in Fig. \ref{fig:mse_vs_Ts}(b), a smaller number of OFDM symbols may degrade the velocity estimation performance due to reduced Doppler frequency resolution. However, our proposed D-VQVAE network demonstrates greater robustness against this effect. This improved robustness can be attributed to the D-VQVAE network's ability to jointly extract time-domain features from the sensing signals of distributed receive APs, thereby mitigating the impact of a limited number of OFDM symbols on velocity estimation.

\begin{figure}[t]
    \centering
    \subfigure[]{\includegraphics[width=1.65in]{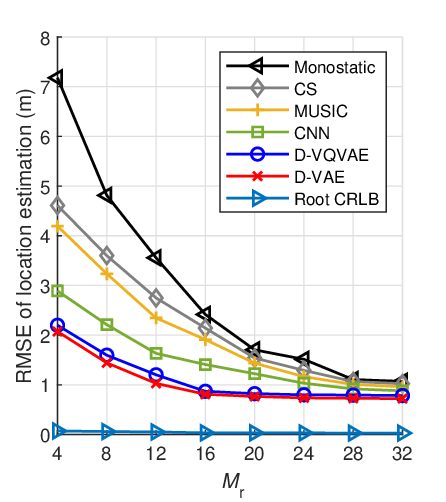}}
    \subfigure[]{
\includegraphics[width=1.65in]{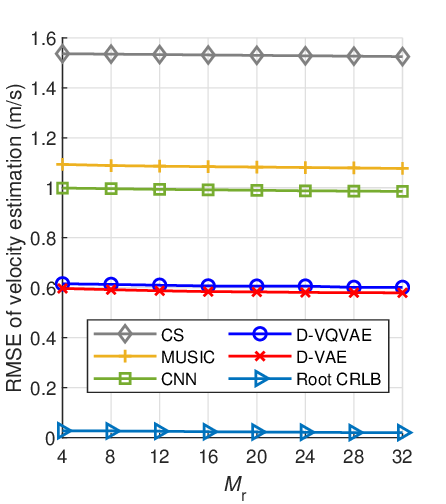}}
        \caption{RMSE for (a) location estimation and (b) velocity estimation versus the number of antennas at the receive AP $M_{\mathrm{r}}$.}
        \label{fig:mse_vs_Mr}
        \vspace{-0. cm}
\end{figure}

In Fig. \ref{fig:mse_vs_Mr}, we study the impact of the number of antennas at each receive AP on the sensing performance. The results indicate that increasing the number of antennas enhances localization performance. This improvement is due to the fact that a larger number of antennas of the receive AP increases the resolution in the spatial domain, allowing the system to better separate the sensing signals corresponding to different targets and differentiate between targets. On the other hand, the number of receive antennas has little impact on velocity estimation performance. This is because velocity estimation primarily relies on the Doppler shift information, which does not benefit significantly from the increased spatial resolution provided by additional antennas.

\begin{figure}[t]
    \centering
\includegraphics[width=2.0in]{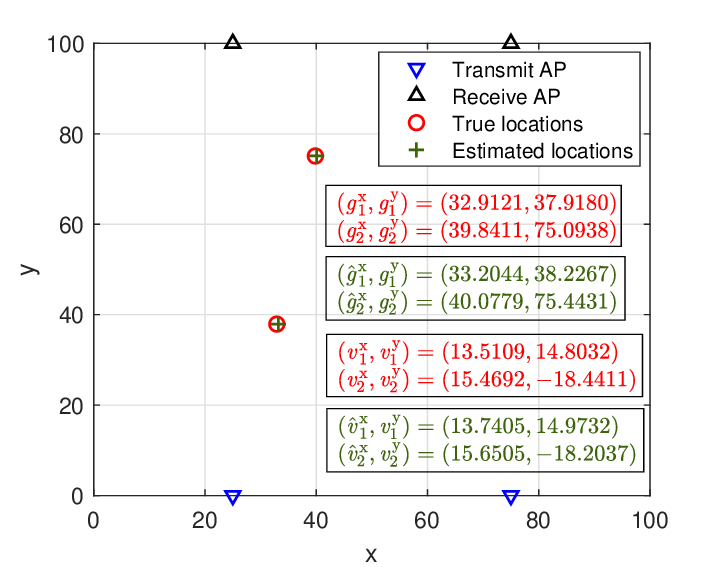}

        \caption{Comparison between the ground truth and the estimated location and velocity ($Q=2$).}
        \label{fig:pos_vel_show}
        \vspace{-0. cm}
\end{figure}

\begin{figure}[t]
        \centering
\includegraphics[width=2.0in]{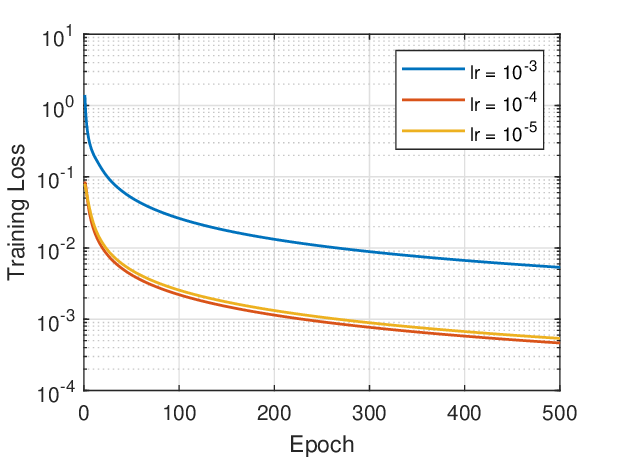}

        \caption{Training loss versus the epoch for different learning rates.}
        \label{fig:convergence}
        \vspace{-0. cm}
\end{figure}

In Fig. \ref{fig:pos_vel_show}, we visualize the results for target localization and velocity estimation and compare
the estimated values with the ground truth. The results in Fig. \ref{fig:pos_vel_show} demonstrate that the proposed D-VQVAE network achieves a high localization accuracy, yielding estimation errors below $1$ m in the considered cell-free MIMO system. Similarly, the estimated velocity of the target is also shown to be close to the ground truth, as can also be observed from the figure.

Next, we study the convergence performance of the proposed schemes in Fig. \ref{fig:convergence}. We evaluate the impact of learning rate (denoted as ``lr'' in Fig. \ref{fig:convergence}) on the convergence performance. The results indicate that the use of a large learning rate (e.g., $10^{-3}$) may not necessarily lead to convergence to a desirable
trained result. On the other hand, a smaller learning rate (e.g., $10^{-4}$ or $10^{-5}$) ensures that the developed D-VQVAE network converges to a satisfactory solution. 
Furthermore, the results indicate that a learning rate of $10^{-4}$  achieves slightly lower training loss compared to that of $10^{-5}$. 
Based on these observations, the learning rate is set to $10^{-4}$ for the training of the D-VQVAE network.

\subsection{Signaling Overhead and Online Execution Runtime}

In this subsection, we analyze the signaling overhead introduced in the fronthaul link for the considered schemes. The analysis is based on the network parameters provided in Table \ref{tb:net}. 
According to the discussion in Section \ref{sec:overhead}, the fully distributed sensing scheme requires 
$96Q = 192$ bits for each receive AP to forward the estimated sensing parameters to the CPU via the fronthaul link. For centralized sensing, the fronthaul signaling overhead is significantly higher, which is given by $2\times 32 \times M_{\mathrm{r}}N_{\mathrm{s}}T_{\mathrm{s}} = 67$ Mbit per update.

\begin{figure}[t]
    \centering
\includegraphics[width=1.8in]{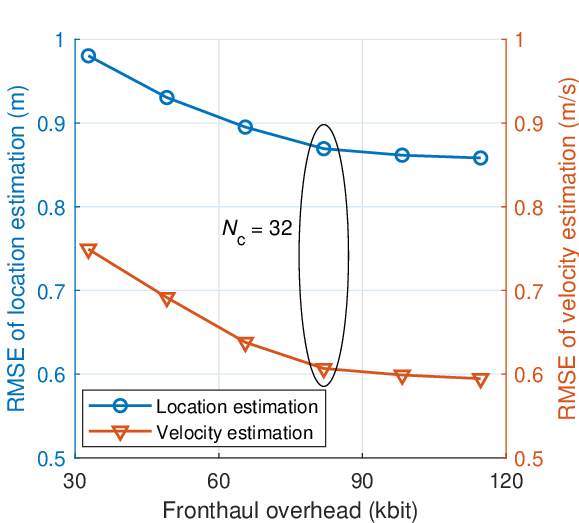}

        \caption{RMSE for location and velocity estimation versus the fronthaul overhead.}
        \label{fig:mse_vs_Nc}
        \vspace{-0. cm}
\end{figure}

In our proposed D-VQVAE network, the signaling overhead is reduced by downsampling the original sensing signals by a factor of $4$ in each dimension. This process significantly reduces the number of feature vectors, resulting in $L = \bar{M}\bar{N}\bar{T} =$ 16,384. Each feature vector is then quantized using a codebook, where each vector is indexed by its corresponding codeword. In Fig. \ref{fig:mse_vs_Nc}, we show the RMSE for location and velocity estimation for varying signaling cost on each fronthaul link. Note that the fronthaul overhead depends on the codebook size $N_\mathrm{c}$. The results demonstrate that a larger codebook reduces the estimation error, as increasing the codebook size enables more precise quantization of the latent representations, allowing the discrete latent features to better capture the useful information from the sensing signals. 
It can be observed that the proposed D-VQVAE network achieves good performance when $N_{\mathrm{c}}=32$, corresponding to a $N_{\mathrm{b}} = 5$-bit codebook. Increasing $N_{\mathrm{c}}$ further provides small performance improvements but may introduce additional complexity to the system. 
When $N_{\mathrm{b}} = 5$, the fronthaul signaling overhead is given by $N_{\mathrm{b}}L = 82$ kbit. Compared with the centralized sensing scheme which incurs $67$ Mbit of overhead, the proposed scheme reduces the signaling overhead on the fronthaul link by $99\%$. Meanwhile, the transmission from each receive AP to the CPU contains essential sensing information, ensuring effective target sensing.

\begin{table}[t]
\centering
\scriptsize{
\caption{Online Execution Runtime}
{\renewcommand{\arraystretch}{1.0}
\begin{tabular}{|c|c|c|c|}
\hline
Settings  & Encoder & Quantization & Decoder  \\ \hline
\begin{tabular}[c]{@{}c@{}} $N_{\mathrm{t}} = N_{\mathrm{r}} = 16$\\ $N_{\mathrm{s}} = T_{\mathrm{s}} = 256$\end{tabular} & $6.75$ ms & $2.51$ ms  & $23.89$ ms \\ \hline
\begin{tabular}[c]{@{}c@{}}$N_{\mathrm{t}} = N_{\mathrm{r}} = 8$\\ $N_{\mathrm{s}} = T_{\mathrm{s}} = 256$\end{tabular} & $2.25$ ms     & $1.34$ ms      & $5.03$ ms      \\ \hline
\begin{tabular}[c]{@{}c@{}} $N_{\mathrm{t}} = N_{\mathrm{r}} = 8$\\ $N_{\mathrm{s}} = T_{\mathrm{s}} = 128$\end{tabular} & $1.59$ ms     & $1.15$ ms     & $2.42$ ms      \\ \hline 
\end{tabular}\label{tb:runtime}}}
\end{table}

Then, in Table \ref{tb:runtime}, we evaluate the online execution runtime for the proposed D-VQVAE network with batch size $B = 20$. We conducted the simulations using a computing server with an Intel Core i7-10700 @ 3.80 GHz CPU and 
an NVIDIA GeForce RTX 3070 GPU. We show the computational time for the encoder, codebook-based quantization, and decoder under various system configurations. The results in Table \ref{tb:runtime} show that the encoding and codebook-based quantization can be completed in only a few milliseconds, demonstrating the minimal computational load on each receive AP and the overall efficiency of our model.

\begin{figure}[t]
    \centering
    \subfigure[]{\includegraphics[width=1.55in]{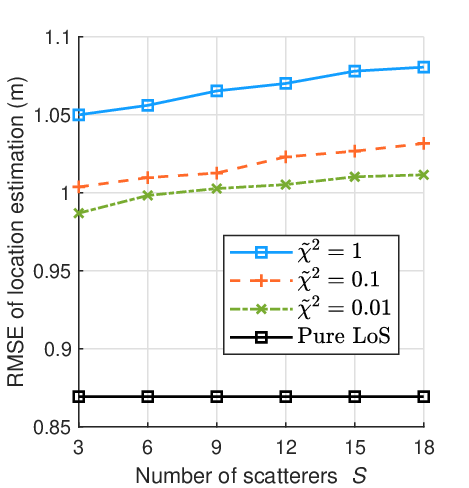}}
    \subfigure[]{
\includegraphics[width=1.55in]{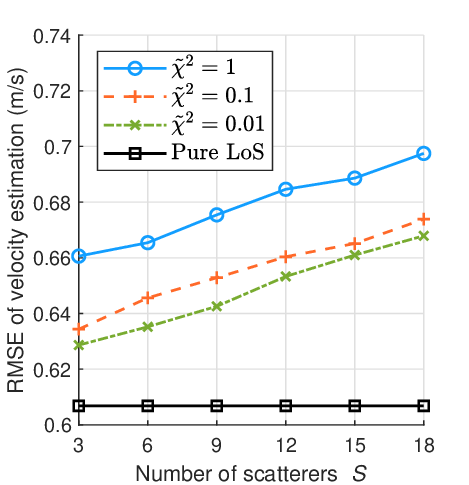}}

        \caption{RMSE for (a) location estimation and (b) velocity estimation versus the number of scatterers $S$.}
        \label{fig:mse_vs_S}
        \vspace{-0.1 in}
\end{figure}

\begin{figure}[t]
    \centering
    \subfigure[]{\includegraphics[width=1.55in]{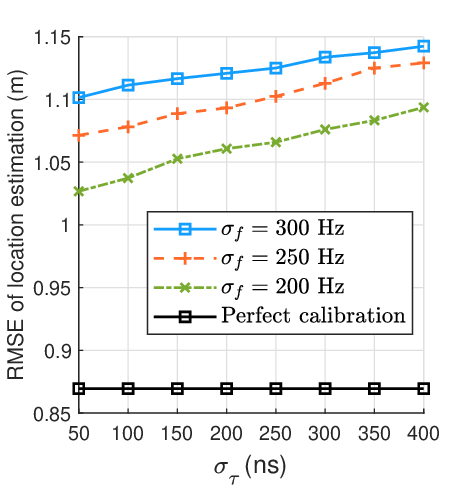}}
    \subfigure[]{
\includegraphics[width=1.55in]{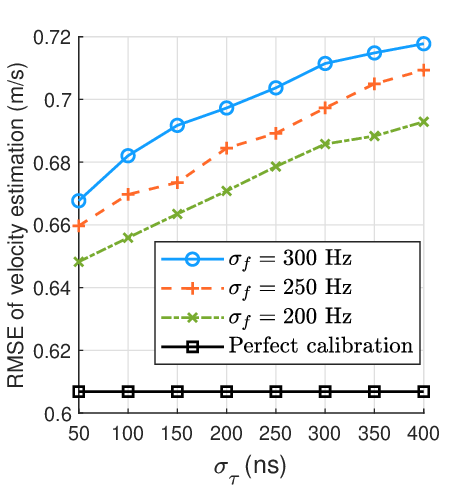}}

        \caption{RMSE for (a) location estimation and (b) velocity estimation under imperfect calibration.}
        \label{fig:mse_asyn}
        \vspace{-0.1 in}
\end{figure} 

\begin{figure*}[b]
\hrule
\begin{align} 
& \hspace{-0.2 in} \frac{\partial\boldsymbol\rho_{i,m}[t]}{\partial g_q^{\mathrm{x}}} = \sum_{n=1}^N \bigg[\frac{\partial A_{n,m,q}}{\partial g_q^{\mathrm{x}}}e^{-j\phi_{i,n,m,q}[t]} + A_{n,m,q}\Big(-j \frac{\partial \phi_{i,n,m,q}[t]}{\partial g_q^{\mathrm{x}}} \Big)e^{-j\phi_{i,n,m,q}[t]}\bigg]\bfa_{\mathrm{r}}( \vartheta_{m,q})\bfa_{\mathrm{t}}^\mathrm{H}(\theta_{n,q}) \mathbf{x}_{i,n}[t] \nonumber \\
&\hspace{0.5 in}+ \sum_{n=1}^N A_{n,m,q}e^{-j\phi_{i,n,m,q}[t]}\bigg[ \frac{\partial \bfa_{\mathrm{r}}( \vartheta_{m,q})}{\partial \vartheta_{m,q}} \frac{\partial \vartheta_{m,q}}{\partial g^{\mathrm{x}}_{q}}\bfa^{\mathrm{H}}_{\mathrm{t}}( \theta_{n,q}) +  \bfa_{\mathrm{r}}( \vartheta_{n,q}) \frac{\partial \bfa_{\mathrm{t}}( \theta_{n,q})}{\partial \theta_{n,q}} \frac{\partial \theta_{n,q}}{\partial g^{\mathrm{x}}_{q}} \bigg] \mathbf{x}_{i,n}[t]. \tag{32} \label{eq:gx derivative}
\end{align}
\end{figure*}

\vspace{-0.0 in}
\subsection{Effect of Multipath Components} \label{sec:multipath simulation}

In this subsection, we evaluate the impact of multipath components on sensing performance. In addition to the LoS sensing channel $\mathbf{G}_{i,n,m}[t]$  shown in (\ref{eq:echo}), we also include the non-LoS component caused by $S$ clutter scatterers, which is modeled as:\bea &&\hspace{-0.4 in}
\tilde{\mathbf{G}}_{i,n,m}[t] = \sum_{s=1}^S \tilde{\beta}_{n,m,s}\sqrt{\mathrm{PL}(\tilde{d}_{n,m,s})} \nonumber \\ 
&& \hspace{-0.1 in} \times e^{-j2\pi (i \tilde{\tau}_{n,m,s}\Delta f-t \tilde{f}_{\mathrm{D},n,m,s}\Delta T)}  
 \bfa_{\mathrm{r}}( \tilde{\vartheta}_{m,s})\bfa_{\mathrm{t}}^\mathrm{H}(\tilde{\theta}_{n,s}),   
\eea 
where $\tilde{\beta}_{n,m,s}\sim \mathcal{CN}(0, \tilde{\chi}^2)$ denotes the complex reflection coefficient of the $s$-th non-LoS scattering path between the $n$-th transmit AP and the $m$-th receive AP.  $\tilde{d}_{n,m,s}$, $\tilde{\tau}_{n,m,s}$, and $\tilde{f}_{\mathrm{D}, n, m, s}$ denote the experienced distance, delay, and Doppler shift through the $s$-th scatterer, respectively.  $\tilde{\theta}_{n,s}$ and $\tilde{\vartheta}_{m,s}$ represent the AoD and AoA of the $s$-th non-LoS path, respectively. We assume the scatterers are randomly distributed within the area. In Fig. \ref{fig:mse_vs_S}, we illustrate the sensing performance for different clutter variances,   $\tilde{\chi}^2 \in \{1, 0.1, 0.01\}$,  and different numbers of scatterers, $S \in \{3, 6, 9, 12, 15, 18\}$. We train the D-VQVAE network on both LoS channels and on channels with varying clutter interference levels. It can be observed that as $S$ increases, the RMSE gradually increases, since the additional scatterers introduce more clutter interference and degrade the sensing signal-to-noise ratio (SNR). Moreover, we notice that under high clutter variance ($\tilde{\chi}^2 = 1$), our model can still yield a satisfactory sensing performance, which demonstrates its
ability to tackle such uncertainties and enable robust sensing.

\subsection{Impact of Clock Asynchronism}

Asynchronous local oscillators at the distributed APs introduce clock asynchronism, which can cause timing offset (TO) and carrier frequency offset (CFO) \cite{XC-TCOM24}. Extensive research has been devoted to time-frequency calibration techniques to estimate and mitigate these offsets. However, completely eliminating TO and CFO is challenging, and small residual offsets usually remain even after calibration. 
In this subsection, we evaluate their impact on the RMSE of location and velocity estimation. Let $\tau_{\mathrm{o}, n,m,t}\sim\mathcal{CN}(0, \sigma^2_{\tau})$ and $f_{\mathrm{o}, n,m,t}\sim\mathcal{CN}(0, \sigma^2_{f})$ denote the residual TO and CFO between the $n$-th transmit AP and the $m$-th receive AP during the $t$-th OFDM symbol interval, respectively. The sensing channel with both TO and CFO can be expressed as follows \cite{XC-TCOM24}:\bea &&\hspace{-0.6 in}
\bar{\mathbf{G}}_{i,n,m}[t] = e^{-j2\pi i  \tau_{\mathrm{o}, n,m,t}\Delta f} e^{j2 \pi f_{\mathrm{o}, n,m,t} t \Delta T} \nonumber \\ &&\hspace{0.2 in}\times \big(\mathbf{G}_{i,n,m}[t] + \tilde{\mathbf{G}}_{i,n,m}[t]\big).  
\eea 
In Fig. \ref{fig:mse_asyn}, we evaluate the impact of clock asynchronism on the sensing performance. The  D-VQVAE network is trained in a dynamic setting that includes LoS channels, multipath with varying clutter levels, and APs exhibiting different timing and frequency offsets. The results show that the proposed network remains robust to the phase errors introduced by unsynchronized clocks. This is because in ISAC-enabled sensing, target range and velocity appear as linear phase slopes across OFDM subcarriers and successive symbols. Although clock asynchronism introduces random phase shifts at each AP and perturbs the phase of the observed echoes, the underlying delay and Doppler information can still be captured by analyzing the slope patterns when given a sufficient number of OFDM symbols. Thus, the overall impact of TO/CFO on localization and velocity accuracy is small, and the proposed D-VQVAE network can still yield reliable estimates.

\vspace{0.2 in}
\section{Conclusion \label{sec:conclusion}}

In this paper, we investigated cooperative ISAC-assisted target sensing in cell-free MIMO systems. Instead of transmitting high-dimensional raw sensing signals from each receive AP to the CPU, we proposed a collaborative processing scheme to split the target sensing procedure between the receive APs and the CPU. To achieve this, we developed a D-VQVAE network, which consists of distributed encoders and codebooks at the receive APs and a decoder at the CPU. The received sensing signals are first encoded by the receive APs locally, followed by codebook-based quantization. Only the indices of the selected codewords are forwarded to the CPU which ensures low signaling overhead on the fronthaul links while providing sufficient sensing information. 
Our simulation results demonstrate that our model outperforms existing baseline schemes and can reduce the signaling overhead by $99\%$ when compared with the centralized sensing scheme. Moreover, it exhibits higher robustness to varying numbers of targets being sensed, ensuring reliable performance in more complex scenarios. 
For future work, we will explore joint system‐level optimization, including AP selection, user association, and beamforming design, together with robust target sensing in dynamic environments. Moreover, it is important to develop DNN architectures and training algorithms that can generalize across varying channel conditions, enabling scalable cooperative ISAC deployment.

\begin{appendices}
\section{Partial Derivatives of $\boldsymbol\rho_{i,m}[t]$\label{sec:appendix1}}

Let $\phi_{i,n,m,q}[t] = 2\pi(i\tau_{n,m,q}\Delta f - t f_{\mathrm{D},n,m,q}\Delta T)$ and $A_{n,m,q} = \beta_{n,m,q}\sqrt{\mathrm{PL}(d_{n,m,q})}$. The LoS sensing channel $\mathbf{G}_{i,n,m}[t]$ in (\ref{eq:echo}) can be rewritten as 
\bea 
\mathbf{G}_{i,n,m}[t] = \sum_{q=1}^Q A_{n,m,q} e^{-j\phi_{i,n,m,q}[t]}\bfa_{\mathrm{r}}( \vartheta_{m,q})\bfa_{\mathrm{t}}^\mathrm{H}(\theta_{n,q}). 
\eea 
Then, for each target $q$, the partial derivative of $\boldsymbol\rho_{i,m}[t]$ w.r.t. $g_q^{\mathrm{x}}$  can be expressed as in (\ref{eq:gx derivative}), which is shown at the bottom of this page. Note that $\frac{\partial\boldsymbol\rho_{i,m}[t]}{\partial g_q^{\mathrm{y}}}$ can be obtained in a similar manner.
In (\ref{eq:gx derivative}), the partial derivatives of $A_{n,m,q}$, $\phi_{i,n,m,q}[t]$, and $\mathbf{a}_{\mathrm{t}}(\theta_{n,q})$ w.r.t. $g_q^{\mathrm{x}}$ are given as follows: 
\setcounter{equation}{32}
\bea 
\frac{\partial A_{n,m,q}}{\partial g_q^{\mathrm{x}}} &\hspace{-0.1 in}=& \hspace{-0.1 in}\beta_{n,m,q} \frac{\partial}{\partial g_q^{\mathrm{x}}}\sqrt{\mathrm{PL}(d_{n,m,q})} \nonumber \\
&\hspace{-0.1 in}=& \hspace{-0.1 in} A_{n,m,q}\left(-\frac{\zeta}{2d_{n,m,q}}\right)\frac{\partial d_{n,m,q}}{\partial g_q^{\mathrm{x}}}, 
\eea 
where 
\bea 
\frac{\partial d_{n,m,q}}{\partial g_q^{\mathrm{x}}} = \frac{g_q^{\mathrm{x}} - t_n^{\mathrm{x}}}{\|\mathbf{t}_n - \mathbf{g}_q\|} + \frac{g_q^{\mathrm{x}} - r_m^{\mathrm{x}}}{\|\mathbf{r}_m - \mathbf{g}_q\|}, 
\eea 
\bea 
\frac{\partial \phi_{i,n,m,q}}{\partial g_q^{\mathrm{x}}} &\hspace{-0.1 in}=& \hspace{-0.1 in}2\pi\left(i\Delta f \frac{\partial \tau_{n,m,q}}{\partial g_q^{\mathrm{x}}} - t \Delta T \frac{\partial f_{\mathrm{D},n,m,q}}{\partial g_q^{\mathrm{x}}} \right) \nonumber \\
&\hspace{-0.1 in}=& \hspace{-0.1 in} \frac{2\pi i \Delta f}{c}\frac{\partial d_{n,m,q}}{\partial g_q^{\mathrm{x}}}, 
\eea
\be 
\frac{\partial \mathbf{a}_{\mathrm{t}}(\theta_{n,q})}{\partial \theta_{n,q}} \hspace{-0.03 in}=\hspace{-0.03 in} j \frac{2\pi d_{\mathrm{t}}}{\lambda_{\mathrm{c}}}\mathrm{diag}\left([0, \ldots, N_{\mathrm{t}}-1]^{\mathrm{T}}\right)\mathrm{sin}(\theta_{n,q})\mathbf{a}_{\mathrm{t}}(\theta_{n,q}), \label{eq:at derivative}
\ee 
and 
\be \frac{\partial \theta_{n,q}}{\partial g_q^{\mathrm{x}}} = -\frac{t_n^{\mathrm{x}} - g_q^{\mathrm{x}}}{d_{n,q}\sqrt{1-\left(\frac{t_n^{\mathrm{x}}-g_q^{\mathrm{x}}}{d_{n,q}}\right)^2}}. \label{eq:theta derivative}
\ee
Note that the partial derivative of $\mathbf{a}_{\mathrm{r}}(\vartheta_{m,q})$  w.r.t. $g_q^{\mathrm{x}}$ can be obtained in a similar manner as in (\ref{eq:at derivative}) and (\ref{eq:theta derivative}). 
Regarding the partial derivative of $\boldsymbol\rho_{i,m}[t]$ w.r.t. the velocity components, we obtain 
\begin{align}
& \hspace{-0.3 in} \frac{\partial\boldsymbol\rho_{i,m}[t]}{\partial v_q^{\mathrm{x}}} = \sum_{n=1}^N A_{n,m,q}\Big(-j \frac{\partial\phi_{i,n,m,q}[t]}{\partial v_q^{\mathrm{x}}}\Big)\nonumber \\ 
&\hspace{0.3 in} \times e^{-j\phi_{i,n,m,q}[t]} \bfa_{\mathrm{r}}( \vartheta_{m,q})\bfa_{\mathrm{t}}^\mathrm{H}(\theta_{n,q})\mathbf{x}_{i,n}[t], 
\end{align} 
where 
\bea 
\frac{\partial\phi_{i,n,m,q}[t]}{\partial v_q^{\mathrm{x}}} &\hspace{-0.12 in}=&\hspace{-0.12 in} -2\pi t \Delta T \frac{\partial f_{\mathrm{D},n,m,q}}{\partial v_q^{\mathrm{x}}} \\ 
&\hspace{-0.12 in}=&\hspace{-0.12 in} -2\pi t \Delta T \frac{f_{\mathrm{c}}}{c}(-\mathrm{cos}(\theta_{n,q})+ \mathrm{cos}(\vartheta_{m,q})). 
\eea 
The partial derivative of $\boldsymbol\rho_{i,m}[t]$ w.r.t. $v_q^{\mathrm{y}}$ can be obtained in a similar manner. 
Finally, the partial derivative of $\boldsymbol\rho_{i,m}[t]$ w.r.t. $\beta_{n,m,q}$ is given by
\bea
&&\hspace{-0.2 in}\frac{\partial\boldsymbol\rho_{i,m}[t]}{\partial\beta_{n,m,q}}
=\sqrt{\mathrm{PL}(d_{n,m,q})}
e^{-j\phi_{i,n,m,q}[t]} \nonumber \\ 
&&\hspace{0.8 in} \times \hspace{0.1 cm}
\mathbf a_{r}(\vartheta_{m,q})
\mathbf a_{t}^{\mathrm{H}}(\theta_{n,q})
\mathbf x_{i,n}[t].
\eea
 
\end{appendices}
\vspace{-0.2 in}


\bibliographystyle{IEEEtran}
\bibliography{refs}

\begin{IEEEbiography}[{\includegraphics[width=1in,height=1.25in,clip,keepaspectratio]{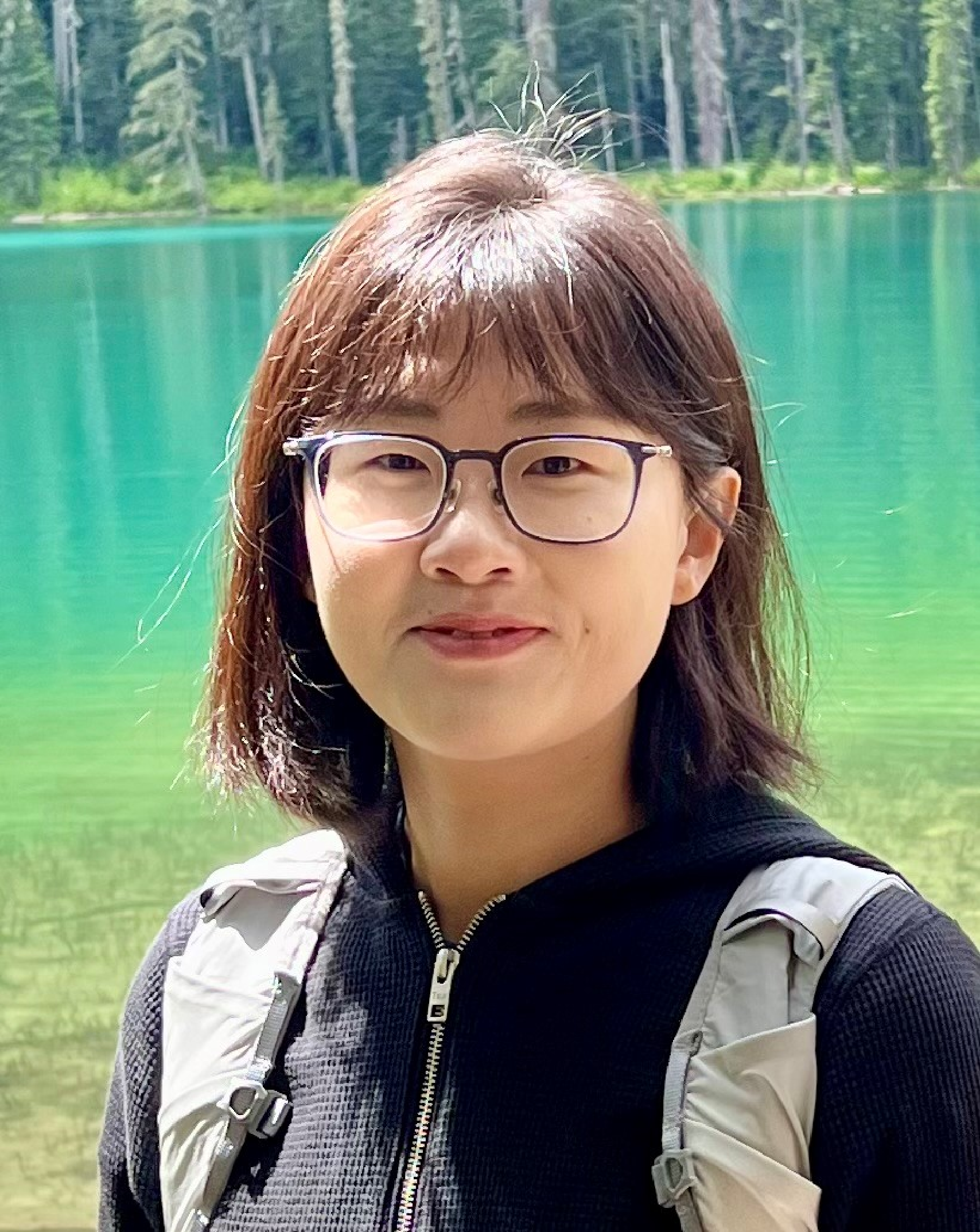}}]
{Zihuan Wang}(Member, IEEE) received the Ph.D. degree from the University of British
Columbia (UBC), Vancouver, Canada, in 2025. She
is currently a postdoctoral fellow in the Electrical and
Computer Engineering Department at the University
of Toronto, Canada. Her research interests include machine learning and optimization for wireless networks, with a main focus on integrated sensing and communication systems and millimeter-wave MIMO systems. She currently serves as the Assistant to Editor-in-Chief of {\it IEEE Transactions on Wireless Communications}. 
She received UBC’s Four Year Fellowship (2020-2024), Li Tze Fong Memorial Fellowship (2023-2024), and the Graduate Support Initiative Award (2021-2023) from the Faculty of Applied Science at UBC. She received the Best Paper Award at the {\it IEEE ICC} 2022. 

\end{IEEEbiography}

\begin{IEEEbiography}[{\includegraphics[width=1in,height=1.25in,clip,keepaspectratio]{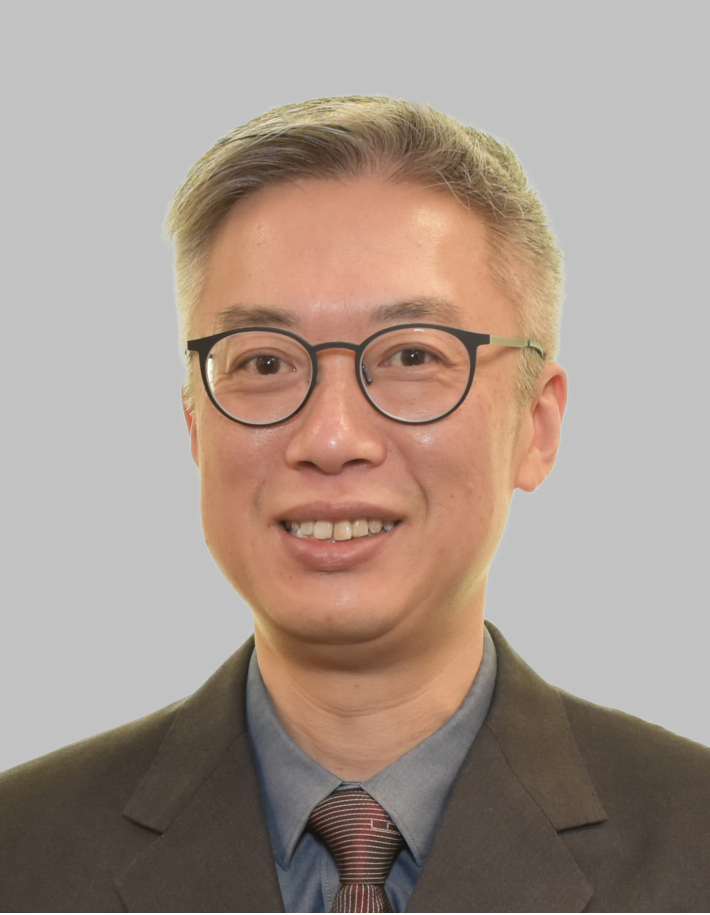}}]{Vincent\;W.S.\;Wong} (Fellow, IEEE) received the B.Sc. degree from the University of Manitoba, Canada, in 1994, the M.A.Sc. degree from the University of Waterloo, Canada, in 1996, and the Ph.D. degree from the University of British Columbia (UBC), Vancouver, Canada, in 2000. From 2000 to 2001, he worked as a systems engineer at PMC-Sierra Inc. (now Microchip Technology Inc.). He joined the Department of Electrical and Computer Engineering at UBC in 2002 and is currently a Professor. His research areas include protocol design, optimization, and resource management of communication networks, with applications to 5G/6G wireless networks, Internet of things, mobile edge computing, smart grid, and energy systems. Dr. Wong is the Editor-in-Chief of the {\it IEEE Transactions on Wireless Communications}. He has served as an Area Editor of the {\it IEEE Transactions on Communications} and {\it IEEE Open Journal of the Communications Society}, an Associate Editor of the {\it IEEE Transactions on Mobile Computing} and {\it IEEE Transactions on Vehicular Technology}, and a Guest Editor of the {\it IEEE Journal on Selected Areas in Communications}, {\it IEEE Internet of Things Journal}, and {\it IEEE Wireless Communications}. Dr. Wong was the General Co-Chair of {\it IEEE INFOCOM} 2024; Tutorial Co-Chair of {\it IEEE GLOBECOM} 2018; Technical Program Co-Chair of {\it IEEE  VTC}2020{\it -Fall} and {\it IEEE SmartGridComm} 2014; and Symposium Co-Chair of {\it IEEE ICC}'18, {\it IEEE SmartGridComm} ('13, '17) and {\it IEEE GLOBECOM}'13. He received the 2022 Best Paper Award from {\it IEEE Transactions on Mobile Computing} and Best Paper Awards at the {\it IEEE ICC} 2022 and {\it IEEE GLOBECOM} 2020. He has served as the Chair of the IEEE Vancouver Joint Communications Chapter and IEEE Communications Society Emerging Technical Sub-Committee on Smart Grid Communications. He was an IEEE Communications Society Distinguished Lecturer from 2019 to 2020 and is an IEEE Vehicular Technology Society Distinguished Lecturer for the term of 2023$-$2026. Dr. Wong is a Fellow of the IEEE, Canadian Academy of Engineering, and the Engineering Institute of Canada.
\end{IEEEbiography}

\begin{IEEEbiography}[{\includegraphics[width=1in,height=1.25in,clip,keepaspectratio]{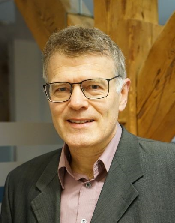}}]{Robert Schober} (Fellow, IEEE) received the Diploma and Ph.D. degrees in electrical engineering from the Friedrich-Alexander University of Erlangen-Nuremberg (FAU), Germany, in 1997 and 2000, respectively. 

From 2002 to 2011, he was a Professor and a Canada Research Chair with the University of British Columbia, Vancouver, Canada. Since January 2012, he has been an Alexander von Humboldt Professor and the Chair for Digital Communication with FAU. His research interests fall into the broad areas of communication theory, wireless and molecular communications, and statistical signal processing.

Prof. Schober received several awards for his work including the 2002 Heinz Maier Leibnitz Award of the German Science Foundation, the 2004 Innovations Award of the Vodafone Foundation for Research in Mobile Communications, the 2006 UBC Killam Research Prize, the 2007 Wilhelm Friedrich Bessel Research Award of the Alexander von Humboldt Foundation, the 2008 Charles McDowell Award for Excellence in Research from UBC, the 2011 Alexander von Humboldt Professorship, the 2012 NSERC E.W.R. Stacie Fellowship, the 2017 Wireless Communications Recognition Award by the IEEE Wireless Communications Technical Committee, the 2022 IEEE Vehicular Technology Society Stuart F. Meyer Memorial Award, and a Honorary Doctorate from Aristotle University of Thessaloniki, Greece, in 2024. Furthermore, he received numerous Best Paper Awards for his work including the 2022 ComSoc Stephen O. Rice Prize and the 2023 ComSoc Leonard G. Abraham Prize. Since 2017, he has been listed as a Highly Cited Researcher by the Web of Science. He is a Fellow of the Canadian Academy of Engineering and the Engineering Institute of Canada, a member of the European Academy of Sciences and Arts and the German National Academy of Science and Engineering. He served as an Editor-in-Chief for the IEEE TRANSACTIONS ON COMMUNICATIONS, VP Publications of the IEEE Communication Society (ComSoc), ComSoc Member at Large, and ComSoc Treasurer. He currently serves as a Senior Editor for Proceedings of the IEEE and as a ComSoc President.
\end{IEEEbiography}

\end{document}